\documentclass[letterpaper]{article}

\usepackage{amsmath} % must load before cleveref
\usepackage{natbib,alifeconf}  %% The order is important
\usepackage{url,hyperref,cleveref}
\usepackage{booktabs}

\usepackage{censor}

\usepackage{mathtools}

\usepackage{multirow}
\usepackage{xcolor} % \crule
\usepackage{amssymb} % for blacksquare and square
\usepackage{adjustbox} % resize tikz
\usepackage{subcaption} % subfigures

\usepackage{wrapfig}
\usepackage{hyperref}

\usepackage[most]{tcolorbox}
\newtcbox{\inlinebox}[1][]{enhanced,
 box align=base,
 nobeforeafter,
 colback=red!5!white,
 colframe=red!75!white,
 size=small,
 left=0pt,
 right=0pt,
 boxsep=2pt,
 arc=0pt
 #1}

\usepackage{tikz-cd}
\usetikzlibrary{backgrounds}

\usepackage{tikz}
\usetikzlibrary{positioning}
\usetikzlibrary{arrows,decorations.markings}

\tikzset{%
  ->-/.style={decoration={markings, mark=at position 0.5 with {\arrow{stealth}}},
              postaction={decorate}}
}

\newcommand{\neckfour}[5]{
\begin{tikzpicture}
      \draw (0,0) circle [radius=#1];
      \foreach \fill [count=\xi, evaluate=\xi as \ang using 360*(\xi/4)] in {#2, #3, #4, #5} {
         \node[circle,draw,fill=\fill,scale=0.8] (\xi) at (\ang:#1){};
      }
\end{tikzpicture}
}

\newcommand\crule[1]{\fcolorbox{black}{#1}{\null}}

\newcommand{\rulethirty}{
\begin{figure}[hbt]
\centering
\begin{adjustbox}{width=\linewidth}
\begin{tikzpicture}[scale=0.50]
    \draw (-0.5,-1.3) rectangle +(4,2.6);
    \filldraw[fill=black] (0,0) rectangle +(0.85,0.85);
    \filldraw[fill=black] (1,0) rectangle +(0.85,0.85);
    \filldraw[fill=black] (2,0) rectangle +(0.85,0.85);
    \filldraw[fill=white] (1,-1) rectangle +(0.85,0.85);

    \draw (3.5,-1.3) rectangle +(4,2.6);
    \filldraw[fill=black] (4,0) rectangle +(0.85,0.85);
    \filldraw[fill=black] (5,0) rectangle +(0.85,0.85);
    \draw (6,0) rectangle +(0.85,0.85);
    \filldraw[fill=white] (5,-1) rectangle +(0.85,0.85);
    
    \draw (7.5,-1.3) rectangle +(4,2.6);
    \filldraw[fill=black] (8,0) rectangle +(0.85,0.85);
    \draw (9,0) rectangle +(0.85,0.85);
    \filldraw[fill=black] (10,0) rectangle +(0.85,0.85);
    \filldraw[fill=white] (9,-1) rectangle +(0.85,0.85);

    \draw (11.5,-1.3) rectangle +(4,2.6);
    \filldraw[fill=black] (12,0) rectangle +(0.85,0.85);
    \draw (13,0) rectangle +(0.85,0.85);
    \draw (14,0) rectangle +(0.85,0.85);
    \filldraw[fill=black] (13,-1) rectangle +(0.85,0.85);
    
    \draw (15.5,-1.3) rectangle +(4,2.6);
    \draw (16,0) rectangle +(0.85,0.85);
    \filldraw[fill=black] (17,0) rectangle +(0.85,0.85);
    \filldraw[fill=black] (18,0) rectangle +(0.85,0.85);
    \filldraw[fill=black] (17,-1) rectangle +(0.85,0.85);

    \draw (19.5,-1.3) rectangle +(4,2.6);
    \draw (20,0) rectangle +(0.85,0.85);
    \filldraw[fill=black] (21,0) rectangle +(0.85,0.85);
    \draw (22,0) rectangle +(0.85,0.85);
    \filldraw[fill=black] (21,-1) rectangle +(0.85,0.85);
    
    \draw (23.5,-1.3) rectangle +(4,2.6);
    \draw (24,0) rectangle +(0.85,0.85);
    \draw (25,0) rectangle +(0.85,0.85);
    \filldraw[fill=black] (26,0) rectangle +(0.85,0.85);
    \filldraw[fill=black] (25,-1) rectangle +(0.85,0.85);
    
    \draw (27.5,-1.3) rectangle +(4,2.6);
    \draw (28,0) rectangle +(0.85,0.85);
    \draw (29,0) rectangle +(0.85,0.85);
    \draw (30,0) rectangle +(0.85,0.85);
    \filldraw[fill=white] (29,-1) rectangle +(0.85,0.85);
\end{tikzpicture}
\end{adjustbox}
\caption{Example of mapping between CA rule and decimal representation. Rule $\square\square\square\blacksquare\blacksquare\blacksquare\blacksquare\square = 000011110_2 = 30_{10}$}
\label{fig:rule30}
\end{figure}
}

\newcommand{\injectivity}{
\begin{figure}[ht]
\centering
\begin{tikzpicture}[scale=0.7]
  \tikzset{triangle arrow/.style={->,>=stealth',thick}}
  \draw (0.2,0) node {$\cdots$};
  \draw (9.8,0) node {$\cdots$};
  \foreach \x in {0,...,4}
    \draw (\x*2 + 1,0) circle (0.3cm);
  \draw[triangle arrow] (1.3,0) -- (2.7,0); % Arrow from first to second circle
  \draw[triangle arrow] (5.3,0) -- (6.7,0); % Arrow from third to fourth circle
  \draw[triangle arrow] (7.3,0) -- (8.7,0); % Arrow from fourth to fifth circle
  \draw[triangle arrow,bend left=45] (3.3,0) to (6.7,0); % Curved arrow from second to fourth circle
  \draw[dotted, thick, gray, ->, >=stealth'] (3.3,0) -- (4.7,0);
\end{tikzpicture}
\end{figure}
}

\title{Networks of Binary Necklaces Induced by Elementary Cellular Automata Rules}

\author{
   Lapo Frati$^{1}$,
   Csenge Petak$^{2}$, \and
   Nick Cheney$^1$ \\
   \mbox{}\\
   $^1$Neurobotics Lab, University of Vermont, USA
   lfrati@uvm.edu\\
   $^2$Biology Department, University of Vermont, USA\\
} % email of corresponding author

\begin{document}
\maketitle

\begin{abstract}
\noindent Elementary cellular automata deterministically map a binary sequence to another using simple local rules. Visualizing the structure of this mapping is difficult because the number of nodes (i.e. possible binary sequences) grows exponentially. If periodic boundary conditions are used, rotation of a sequence and rule application to that sequence commute. This allows us to recover the rotational invariance property of loops and to reduce the number of nodes by only considering binary \textit{necklaces}, the equivalence class of n-character strings taking all rotations as equivalent. Combining together many equivalent histories reveals the general structure of the rule, both visually and computationally. In this work, we investigate the structure of necklace-networks induced by the 256 Elementary Cellular Automata rules and show how their network structure change as the length of necklaces grow.
\end{abstract}

\section{Introduction}
Complex and interesting patterns can emerge from very simple algorithms. Among the most well known and studied models of such emergence of complexity from simple rules are Cellular Automata (CA), first introduced by von Neumann and Ulam for modeling biological self-reproduction \citep{von1966theory}. Elementary CA (ECA) represent a deterministic mapping between 1-D binary strings into other strings by using only local rules. For each position in the string an ECA rule determines what the next binary value should be using only the current value at that position and the value of its immediate neighbors, left and right. Each ECA rule can be represented by just 8 binary values (see Figure \ref{fig:rule30}), one for each of the possible combinations of left, center and right values, for a total of $2^8=256$ rules.
ECA rules are a foundational substrate that can be used as models of physical and biological systems \citep{chopard1998cellular, ermentrout1993cellular} and artificial life \citep{langton1986studying}. Over the years research built on this foundation and used CAs for amazing results in alife such as softbots \citep{cheney2014unshackling} and xenobots \citep{blackiston2021cellular}.
\begin{figure}[ht]
    \centering
    \begin{tikzcd}[show background rectangle]
    {\;\crule{white}\crule{white}\crule{white}\crule{black}\crule{black}} \arrow[r, "rshift"] \arrow[d, "rule"]
    & {\;\crule{black}\crule{white}\crule{white}\crule{white}\crule{black}} \arrow[d, "rule"] \\
    {\;\crule{white}\crule{white}\crule{black}\crule{white}\crule{white}} \arrow[r, "rshift"]
    & {\;\crule{white}\crule{white}\crule{white}\crule{black}\crule{white}}
    \end{tikzcd}
    \caption{Commutativity of shift and rule application under periodic boundary conditions}
    \label{fig:commutativity}
\end{figure}
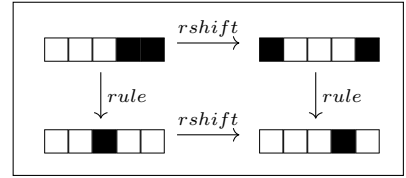
\rulethirty
Fields such as AI, RL, and robotics are fascinated with controlling agents capable of interacting and reacting to complex environments. To effectively cope with dynamic environments agents need to match their internal dynamics to external ones. Rhythms and frequencies can, for example, determine if a gait is suitable for locomotion, or if complex actions are taken at the proper time.
State-of-the-art approaches in these fields employ massive and sophisticated models to learn or evolve the necessary temporal dynamics. Here, instead, we look at the other extreme and ask ourselves what temporal structure is hidden in the complex dynamics of simple systems such as ECA.

The structure of connections between strings determined by a rule can reveal interesting properties of that rule. Each one of the 256 mappings is deterministic, so each possible string has a unique successor but could have zero (not surjective) or multiple predecessors (not injective). The existence of strings without predecessors, called ``Garden of Eden" configurations are linked to the local injectivity of a rule, for example since each state has exactly one successor, having multiple predecessors implies that at least one state won't have predecessors (see \textcolor{gray}{dotted} line below)
\injectivity

Empirically analyzing the structure of networks induced by ECA rules becomes quickly intractable as the number of nodes in such networks grow exponentially, since a binary string of length N has $2^N$ distinct configurations.
However, when using periodic boundary conditions, sequences become a closed binary loop. When dealing with a closed loop, the application of a CA rule is commutative with respect to a right/left shift operation (see \textbf{Figure \ref{fig:commutativity}}).
Therefore, if two states are the same when rotated by X, all their descendants (i.e., following results of rule application) will also be the same if rotated by X, thus preserving the loop structure. Grouping together strings that only differ by a rotation we can remove the redundancy caused by representing loops as strings.
In combinatorics the equivalence class of binary strings of length $N$, taking all rotations as equivalent is called a binary \textit{necklace} (see \textbf{Table \ref{tab:4necklaces}}). 
For brevity we are going to refer to binary necklaces as just necklaces since that's the only kind of necklace considered in this work. From each equivalence class we are going to select the element with the smallest decimal representation (or equivalently with the largest number of zeros on the left side of its binary representation) as the \textit{representative} of that equivalence class.

\subsubsection{In this work we explore:}
\begin{figure}[!htb]
    \centering
    \includegraphics[width=0.8\linewidth]{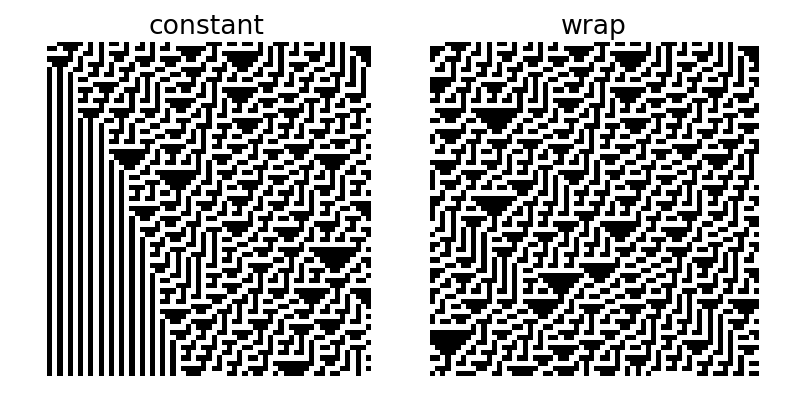}
    \caption{Different boundary conditions for rule 30 starting from a random seed. \textit{Left}: Padding the sequence with a constant value of zero. \textit{Right}: Padding using a periodic boundary condition.}
    \label{fig:constant_vs_wrap}
\end{figure}

\begin{itemize}
    \item the distribution of binary necklaces (Figure \ref{fig:distribution_necklaces}),
    \item network coarse-graining using necklaces (Figures \ref{fig:rule90_rule_to_graph}, \ref{fig:coarse_graining}),
    \item network growth as necklaces' length increases (Figures \ref{fig:rule90_progression} \ref{fig:grid_110}, \ref{fig:grid_45}),
    \item and finally show an overview of necklace-networks for all 256 rules (Figure \ref{fig:largest_components}).
\end{itemize}
Code available at: \url{github.com/lfrati/necklaCA}.
A common alternative to using periodic boundary conditions consists in adding one extra element of padding on both ends of the string before applying the rule, e.g. $\blacksquare\blacksquare\blacksquare \xrightarrow{\text{pad}} \square\blacksquare\blacksquare\blacksquare\square \xrightarrow{\text{rule}} \blacksquare\square\blacksquare$ but it can introduce artifacts in the evolution of a rule (see \textbf{Figure \ref{fig:constant_vs_wrap}}).

\begin{table}[!htb]
    \centering
    \begin{tabular}{c|c}
    \neckfour{0.3}{white}{white}{white}{white}
        & \crule{white}\crule{white}\crule{white}\crule{white}\\
        \\
        \hline
    \multirow{ 4}{*}{\neckfour{0.3}{white}{white}{white}{black}}
        & \crule{black}\crule{white}\crule{white}\crule{white}\\
        & \crule{white}\crule{black}\crule{white}\crule{white}\\
        & \crule{white}\crule{white}\crule{black}\crule{white}\\
        & \crule{white}\crule{white}\crule{white}\crule{black}\\
        \\
        \hline
    \multirow{ 4}{*}{\neckfour{0.3}{white}{white}{black}{black}}
        & \crule{black}\crule{black}\crule{white}\crule{white}\\
        & \crule{white}\crule{black}\crule{black}\crule{white}\\
        & \crule{white}\crule{white}\crule{black}\crule{black}\\
        & \crule{black}\crule{white}\crule{white}\crule{black}\\
    \end{tabular}
    \begin{tabular}{|c|c}
    \multirow{ 4}{*}{\neckfour{0.3}{white}{black}{black}{black}}
        & \crule{black}\crule{black}\crule{black}\crule{white}\\
        & \crule{white}\crule{black}\crule{black}\crule{black}\\
        & \crule{black}\crule{white}\crule{black}\crule{black}\\
        & \crule{black}\crule{black}\crule{white}\crule{black}\\
        \\
        \hline
        \\
    \neckfour{0.3}{black}{black}{black}{black}
        & \crule{black}\crule{black}\crule{black}\crule{black}\\
        \\
        \hline
        \\
    \multirow{ 2}{*}{\neckfour{0.3}{white}{black}{white}{black}}
        & \crule{white}\crule{black}\crule{white}\crule{black}\\
        & \crule{black}\crule{white}\crule{black}\crule{white}\\
    \end{tabular}
    \caption{4-necklaces and corresponding binary sequences.}
    \label{tab:4necklaces}
\end{table}

\section{How many necklaces are there?}
As we can see from \textbf{Table \ref{tab:4necklaces}} many strings can correspond to the same necklace but not all necklaces contain the same number of elements. The ``all white" and ``all black" necklaces only contain a single element, and the ``alternating black white" necklace only contains 2 elements. How does the number of necklaces grow as their length increase? Some examples of the number of n-bead necklaces with 2 colors can be found in the Online Encyclopedia of Integer Sequences \citep{OEIS_A000031}. 

More generally it can be shown that from Pólya's enumeration theorem applied to the action of the cyclic group $C_{n}$ acting on the set of all functions $f:\{1, ..., n\}\rightarrow \{0, 1\}$ follows that the number of unique binary necklaces of length n is:
\begin{equation}
    N_2(n) = \frac{1}{n} \sum_{d|n} \varphi(d) \cdot 2^{n/d}
\end{equation}
where $\varphi(d)$ is Euler's totient function.
\begin{figure}[htb]
    \centering
    \includegraphics[width=\linewidth]{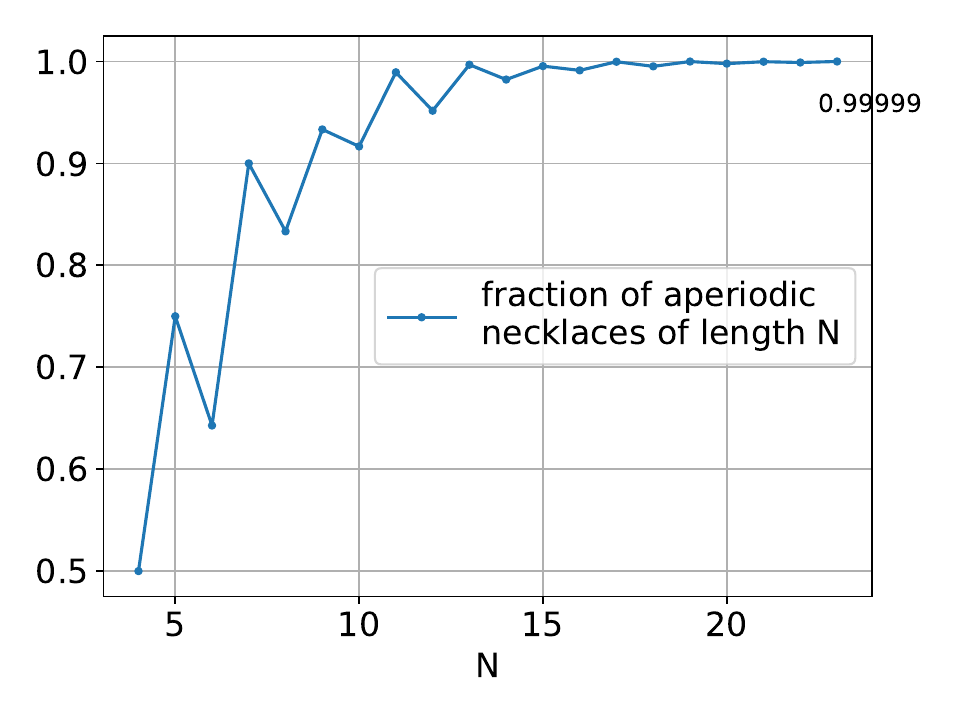}
    \caption{The fraction of aperiodic binary necklaces quickly approaches 1 as the sequence length N increases.}
    \label{fig:aperiodic_necklaces}
\end{figure}
\begin{figure}[htb]
    \centering
    \includegraphics[width=\linewidth]{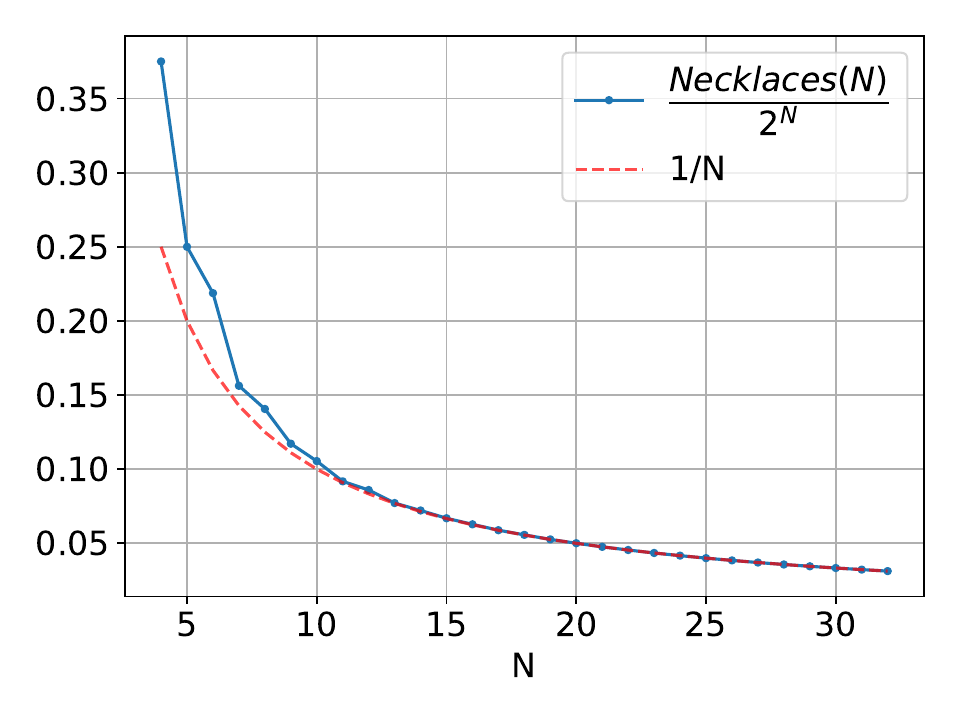}
    \vspace{-0.3in}
    \caption{Scaling of number of necklaces as N grows: While the number of necklaces grows exponentially as the length of necklaces increases it grows only as $2^{N-1}$. So the ratio of necklaces and sequences actually shrinks as $\frac{1}{N}$}
    \label{fig:scaling_necklaces}
\end{figure}
Empirically we observe (see \textbf{Figure \ref{fig:aperiodic_necklaces}}) that the fraction of aperiodic necklaces (i.e. necklaces that contain N elements for a sequence of length N) quickly approaches 1, meaning that as N grows nearly every equivalence class contains N sequences.
Since there are $2^N$ binary sequences of length $N$, if the size of necklaces approaches $N$ the number of distinct necklaces grows approximately as $\frac{2^N}{N}$. This means that the ratio of necklaces to possible sequences shrinks as $\frac{1}{N}$ (see \textbf{Figure \ref{fig:scaling_necklaces}}). Because of this asymptotic behavior, while the total number of necklaces still grows, their effectiveness at capturing the structure of the rules improves as N grows (i.e. only a smaller and smaller fraction of all possible sequences needs to be investigated e.g. for $N=32$ only $\sim3$\% of all possible sequences are unique necklaces).

\section{On the structure and scaling of necklaces}
\begin{figure}[!htb]
    \centering
    \includegraphics[width=\linewidth]{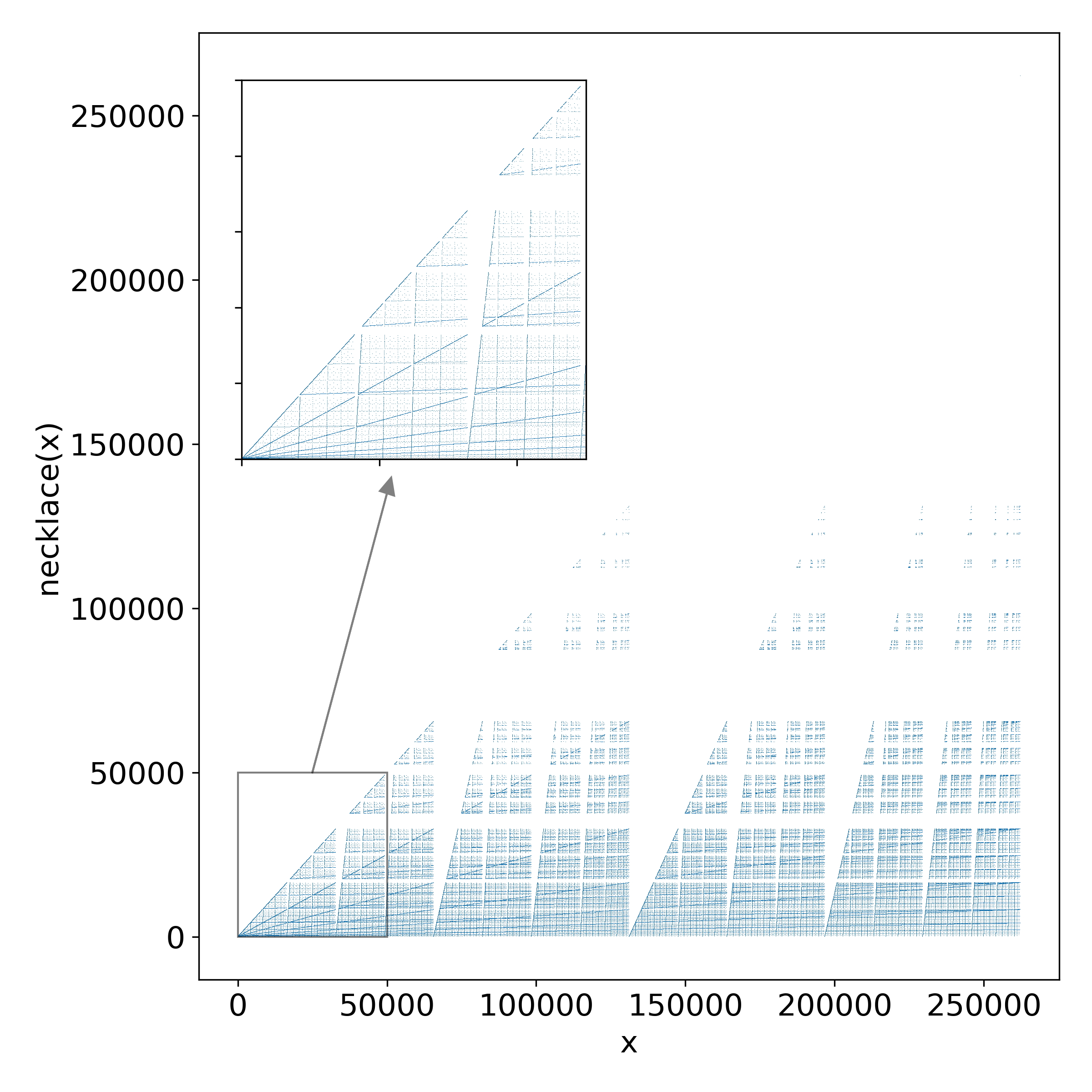}
    \vspace{-0.3in}
    \caption{The structure of necklaces (here $\forall x : x \in [0,...,2^{18}]$) shows self-similarity and several straight lines corresponding to subsets of necklaces with regular scaling.}
    \label{fig:distribution_necklaces}
\end{figure}

In the rest of our work we are going to investigate networks where nodes correspond to necklaces so we first explore the distribution of necklaces on the number line, and how this changes as $N$ increases. In \textbf{Figure \ref{fig:distribution_necklaces}} we plot the value of the representative of that necklace (i.e. the element of that equivalence class with the smallest binary representation) for each sequence $x\in{0, ..., 2^{18}}$. See \textbf{Table \ref{tab:representatives}} for a few examples of the case $N=3$.
Necklaces used in our experiments have been generated with a parallel algorithm using Numpy and Numba \citep{lam2015numba}. This algorithm explores all possible rotations of each sequence of length $N$, up to $N=32$. This approach requires testing $N 2^N$ (a total of $137\,438\,953\,472$ possible sequences for $N=32$) sequences but more efficient algorithms exist that directly generate only the necklaces, such as \cite{fredricksen1986algorithm} or \cite{ruskey2000fast}. Because the number of unique necklaces grows as $\frac{2^N}{N}$ the brute force search incurs in a penalty of $N^2$, which is reduced to $N$ if we need to generate the necklace for each possible value of $\text{x}\in[0,...,2^N]$. 

\begin{table}[htb]
\centering
\begin{tabular}{c|c|c|c}
$\text{x}_{10}$ & $\text{x}_2$ & necklace(x)$_2$ & necklace(x)$_{10}$ \\ \hline
0     & $\square\square\square$ & $\square\square\square$           & 0  \\
1     & $\square\square\blacksquare$ & $\square\square\blacksquare$           & 1 \\
2     & $\square\blacksquare\square$ & $\square\square\blacksquare$           & 1 \\
3     & $\square\blacksquare\blacksquare$ & $\square\blacksquare\blacksquare$      & 3 \\
4     & $\blacksquare\square\square$ & $\square\square\blacksquare$                & 1 \\
5     & $\blacksquare\square\blacksquare$ & $\blacksquare\square\blacksquare$      & 5 \\
6     & $\blacksquare\blacksquare\square$ & $\square\blacksquare\blacksquare$      & 3 \\
7     & $\blacksquare\blacksquare\blacksquare$ & $\blacksquare\blacksquare\blacksquare$ & 7 \\
\end{tabular}
\caption{Mapping between numbers and their necklace representative for $N=3$.}
\label{tab:representatives}
\end{table}

In \textbf{Figure \ref{fig:distribution_necklaces}} we can see the distribution of necklaces on the number line shows remarkable structure, which is even more apparent by comparing the structure of necklaces for consecutive values of $N$. For example, since for each necklace we select as representative the sequence with the smallest decimal representation all the representatives are odd (with the exception of zero).  While the distribution of necklaces shows a self-similar structure across different values of $N$ the actual overlap between the necklaces corresponding to consecutive values of $N$ decreases as the value of $N$ grows (see \textbf{Table \ref{tab:N_to_Nplus_mismatch}}).

\begin{table}[htb]
\centering
\begin{tabular}{l|r|r|r|l}
 & $\text{x}_{10}$ & $\text{x}_2$ & necklace(x)$_2$ & necklace(x)$_{10}$ \\ \hline
N=4 & 9 & $\blacksquare\square\square\blacksquare$ & $\square\square\blacksquare\blacksquare$ & 3 \\
N=5 & 9 & $\square\blacksquare\square\square\blacksquare$ & 
$\square\square\blacksquare\square\blacksquare$ & 5 \\ 
N=4 & 12 & $\blacksquare\blacksquare\square\square$ & $\square\square\blacksquare\blacksquare$ & 3 \\
N=5 & 12 & $\square\blacksquare\blacksquare\square\square$ & $\square\square\square\blacksquare\blacksquare$ & 3 \\
\end{tabular}
\caption{As N increases from 4 to 5, the necklace for 9 changes from 3 to 5, but the necklaces for 12 remain the same.}
\label{tab:N_to_Nplus_mismatch}
\end{table}

In particular we observe that necklaces for $N$ and $N+1$ match for all values $\{ x : 0 \leq x \leq 2^{\lfloor N/2\rfloor + 1}\}$ (see \textbf{Table \ref{tab:divergence_point}}). The behavior past that point is harder to characterize as it depends on where the pair of values at the edge of a sequence falls inside the necklace, but note that the addition of a new $\square$ can not decrease the value of the necklace.

\begin{table}[htb]
\centering
\begin{tabular}{r|r|l}
 x$_2$ & neck$_2$ & neck$_{10}$ \\ \hline
$\underbrace{\square\square}_2\blacksquare\underbrace{\square\square\square\square}_4\blacksquare$   & $\underbrace{\square\square\square\square}_4\blacksquare\underbrace{\square\square}_2\blacksquare$ & 9 \\
$\underbrace{\square\square\square}_3\blacksquare\underbrace{\square\square\square\square}_4\blacksquare$   & $\underbrace{\square\square\square\square}_4\blacksquare\underbrace{\square\square\square}_3\blacksquare$ & 17 \\
\end{tabular}
\caption{When the left most $\blacksquare$ is past the mid point of the sequence, the number of $\square$ on the right can surpass the ones on the left (2 vs 4). When going from $N$ to $N+1$ one more $\square$ is added on the left ($\square\square \rightarrow \square\square\square$) and that changes the necklace ($9 \rightarrow 17$).}
\label{tab:divergence_point}
\end{table}

\section{From rules to networks}
\begin{figure}[!htb]
    \centering
    \includegraphics[width=0.75\linewidth]{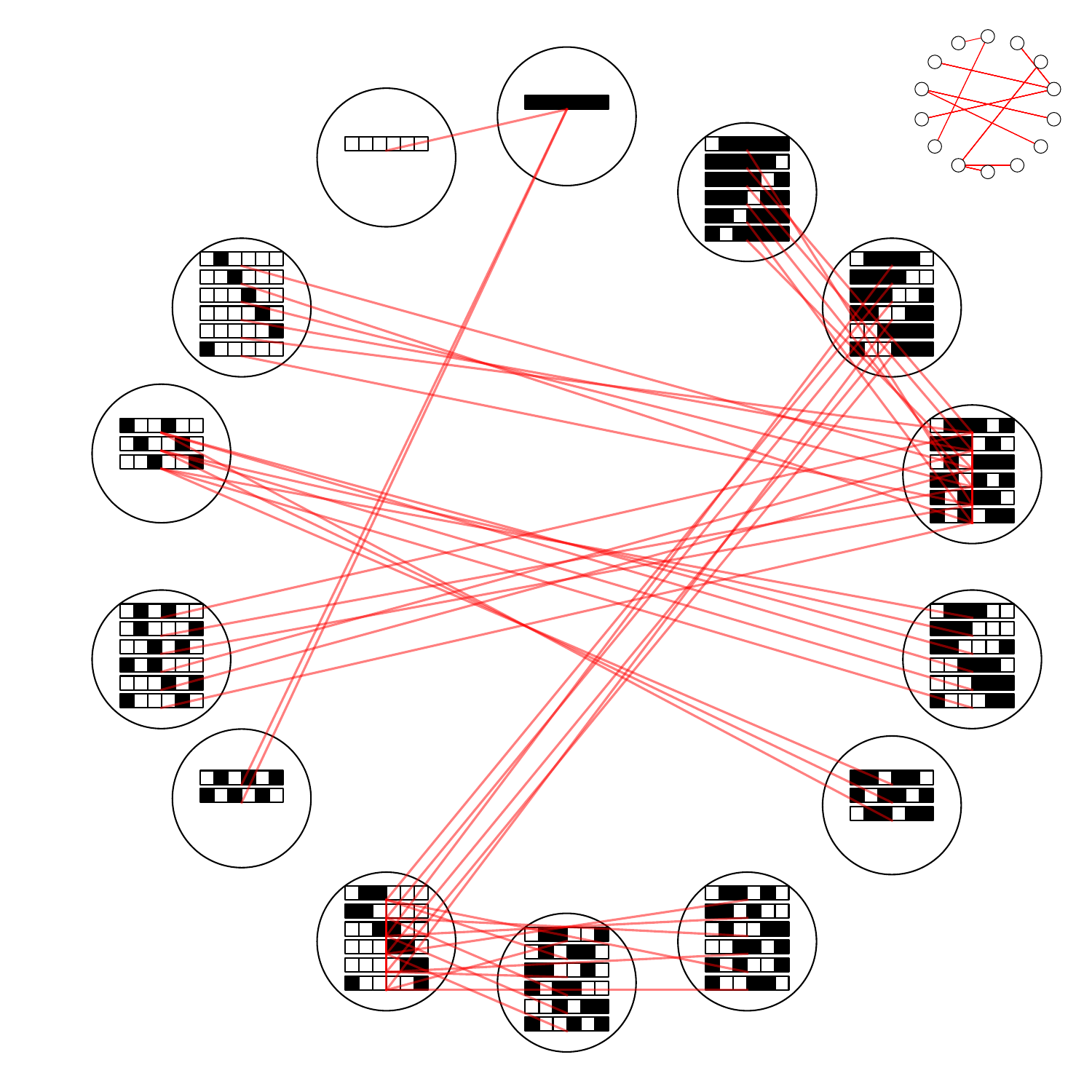}
    \caption{Converting sequence network (bot left) to necklace network (top right) preserves single-successor structure. Rule 90 for $N=6$ is shown.}
    \label{fig:rule90_rule_to_graph}
\end{figure}

As mentioned previously, because the rule application and right/left shift operation commute (see \textbf{Figure \ref{fig:commutativity}}) when considering binary sequences with a periodic boundary we can group states into \textit{necklaces}.
Let $$\text{apply}(\text{rule}:[0,...,255],\text{sequence}) = \text{sequence}'$$ be the result of applying a specific ECA rule to a binary sequence, and $$\text{eq.class}(s)$$ the set of all sequences that belong to the same equivalence class as sequence $s$, then the temporal evolution of a 1-D sequence under rule application will be such that for each sequence $x$ the following holds $$\text{eq.class}(\text{apply}(\text{rule}, x)) = \{\text{apply}(\text{rule}, y) :  y \in \text{eq.class}(x)\}$$
This property is exemplified in \textbf{Figure \ref{fig:rule90_rule_to_graph}}. Because of this property, we can coarse-grain a graph s.t. each node corresponds to an equivalence class (see top-right thumbnail in Figure \ref{fig:rule90_rule_to_graph})

\begin{figure}[!htb]
    \centering
    \includegraphics[width=0.74\linewidth]{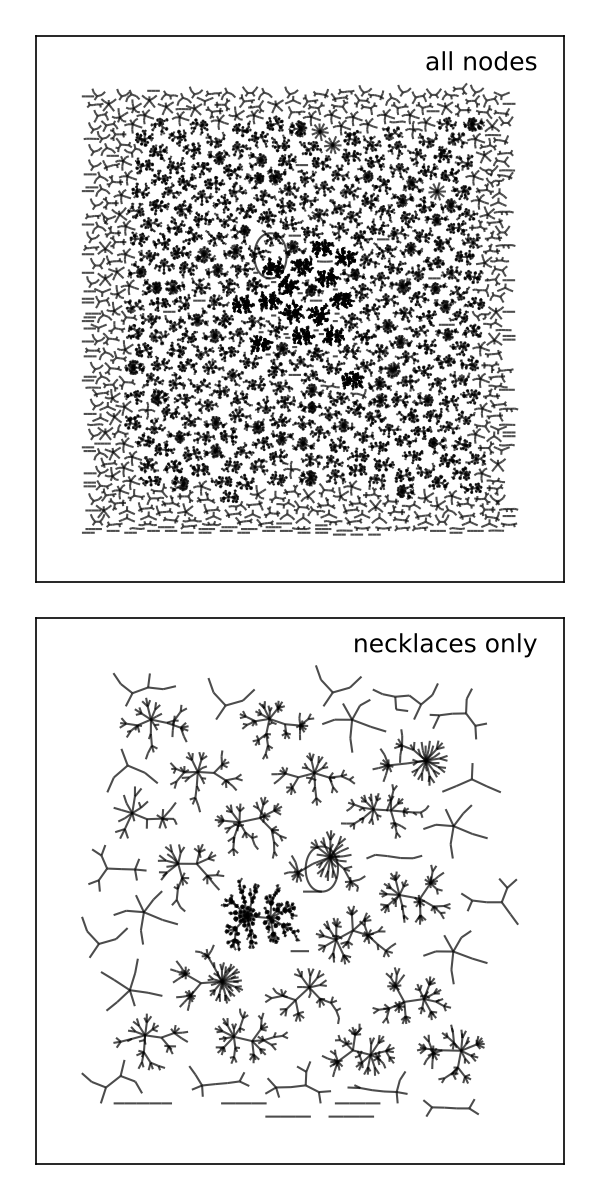}
    \caption{Necklace coarse-graining on rule 179 for N=15 reduces the number of nodes by approx. 15 times.}
    \label{fig:coarse_graining}
\end{figure}

The coarse-graining of networks of necklaces induced by ECA rules allows us to investigate their connectivity structure without wasting resources on redundant paths (i.e. only need to process the representative of each equivalence class, see \textbf{Figure \ref{fig:coarse_graining}}). 

\begin{figure}[!h]
\centering
\begin{subfigure}[b]{\linewidth}
    \centering
    \includegraphics[width=0.85\linewidth]{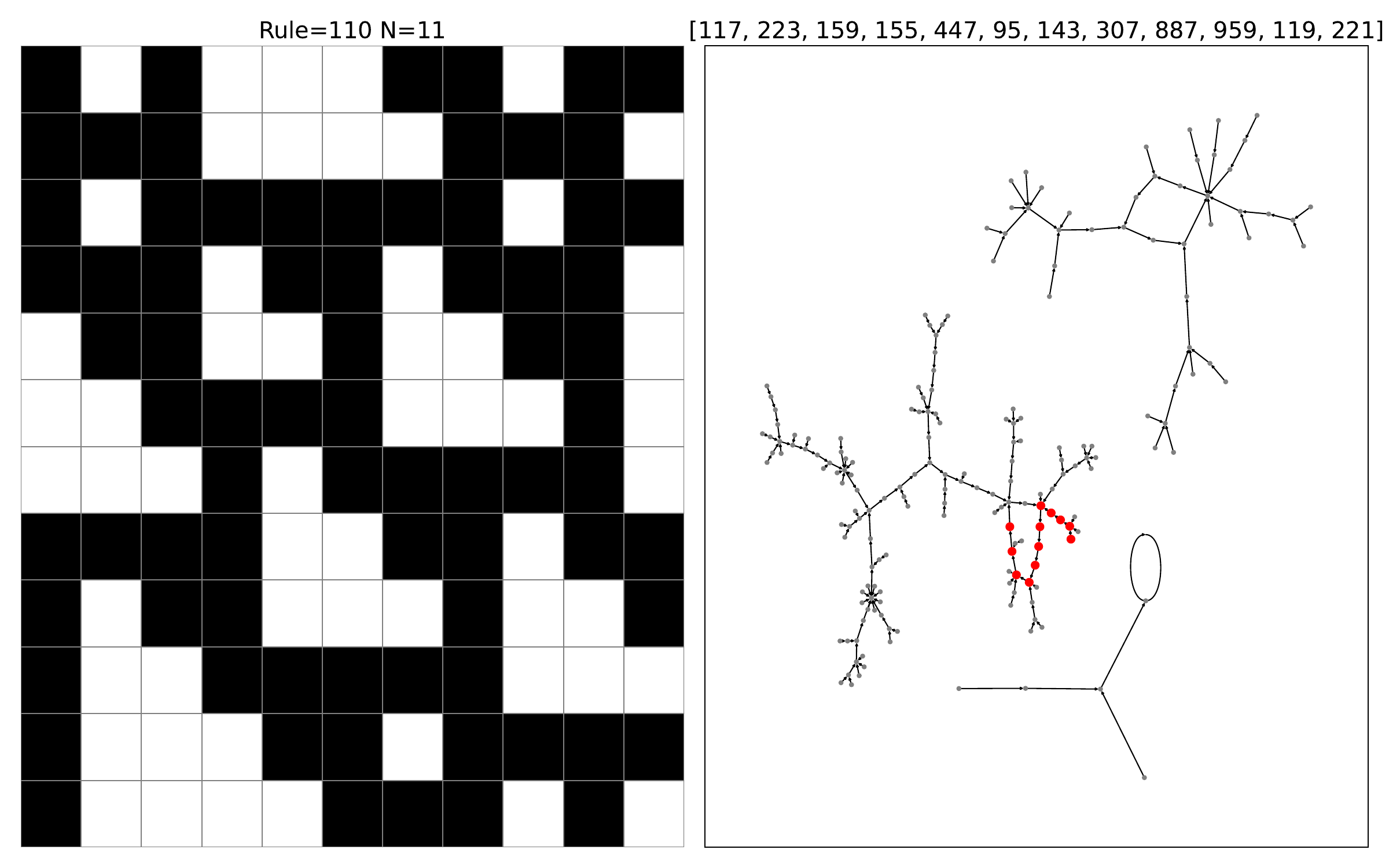}
    \caption{Rule 110, sequence length 11.}
    \label{fig:rule110_seq_and_graph}
\end{subfigure}
\\ \vspace{0.1in}
\begin{subfigure}[b]{\linewidth}
    \centering
    \includegraphics[width=0.85\linewidth]{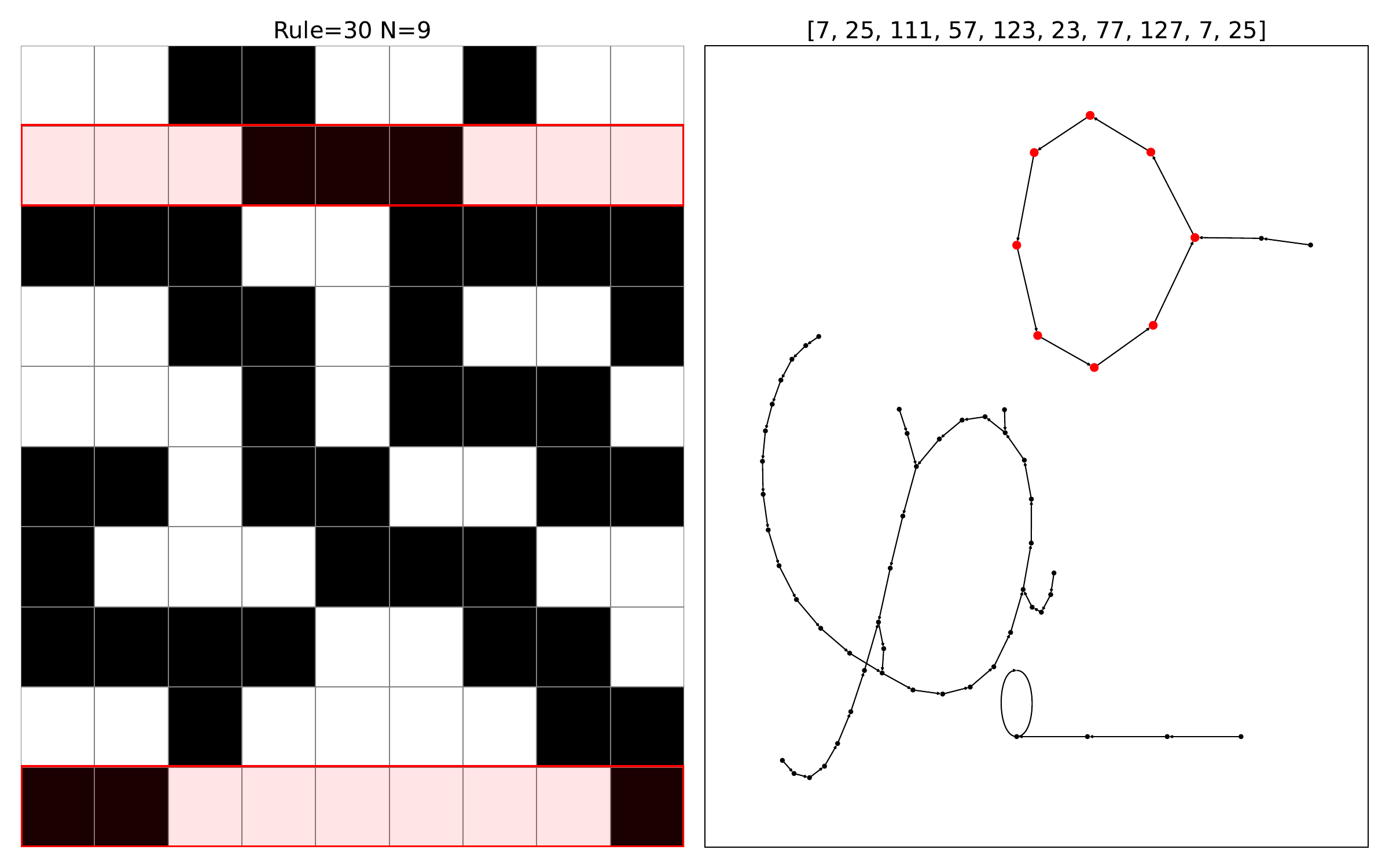}
    \caption{Rule 30, sequence length 9.}
    \label{fig:rule30_loop}
\end{subfigure}
\caption{\textit{Left}: Application of a rule to a random binary string.\;\inlinebox{Boxes} highlight the beginning and end of a loop. \textit{Right}: Necklaces network for the same rule and length.\;\textcolor{red}{Red nodes} highlight the necklaces that sequences on the left belong to. Numbers in square brackets are the decimal value of the representatives of necklaces shown.}
\label{fig:seqs_and_graphs}
\end{figure}

Furthermore, the coarse grained graphs are both easier to visualize (fewer nodes, see \ref{fig:rule110_seq_and_graph}) and easier to interpret (cycles in a graph are easily recognized, see \textbf{Figure \ref{fig:rule30_loop}})

\section{Growing Networks}
While necklaces allow us to simplify the networks induced by ECA rules, their structure changes as the length of sequences considered grows, depending on the rule. In the previous examples we've shown networks generated by sequence of length 9 or 11, as this value strikes an empirical balance between visual richness and readability. 
Furthermore, the networks generated by these rules will likely contain multiple disconnected components. In the following we consider only the Largest Weakly Connected \footnote{In a directed graph a component is weakly connected if there is a path between any two nodes, not necessarily going both ways.} (LWC) component for each network. Networks are shown using the Kamada Kawai \citep{kamada1989algorithm} layout for smaller ones and the Scalable Force-Directed Placement \citep{hu2005efficient} layout for larger ones.
\begin{figure}[htb]
    \centering
    \includegraphics[width=\linewidth]{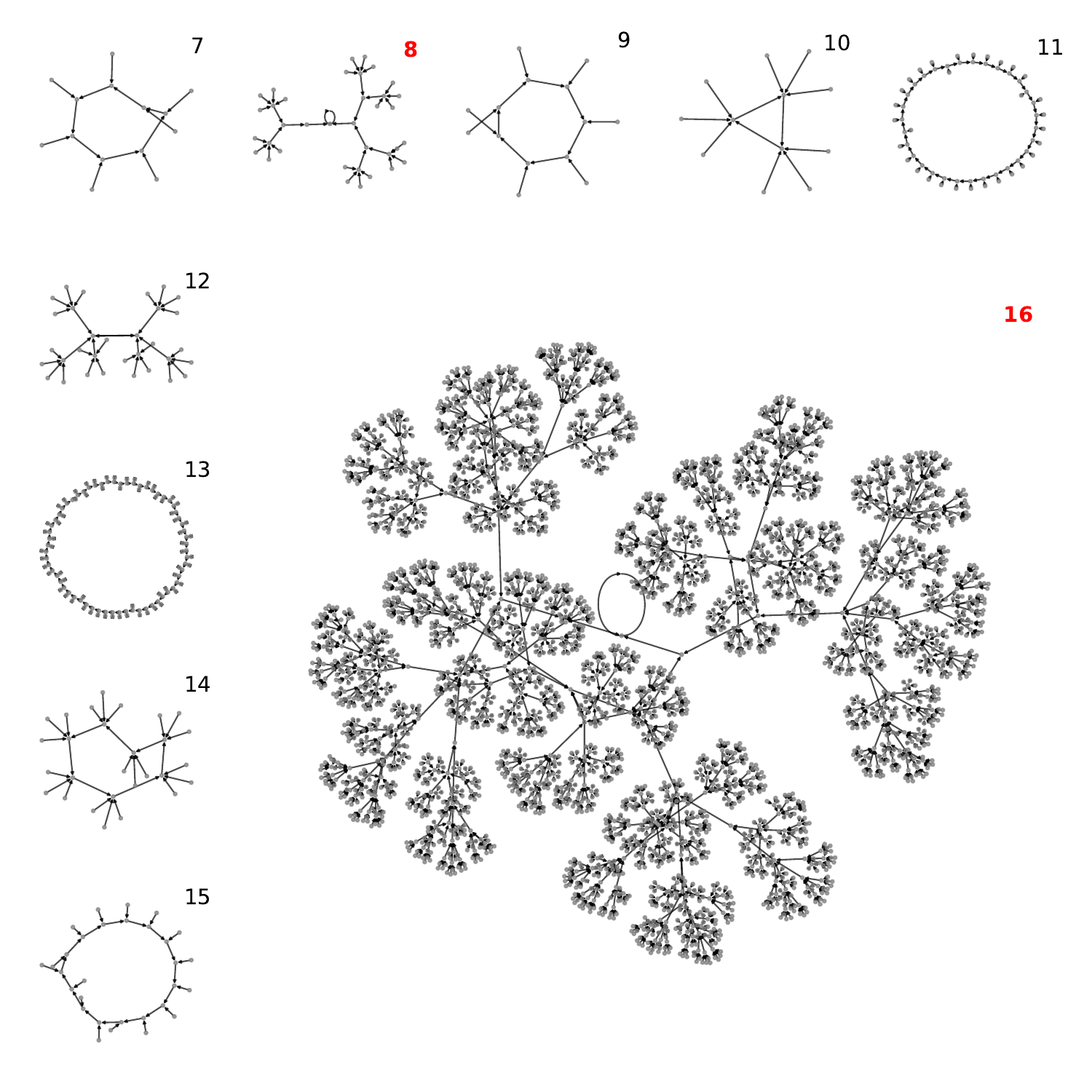}
    \caption{Largest weakly connected components of rule 90 for sequence lengths 7 to 16.\;\textcolor{red}{8} \& \textcolor{red}{16}: rule 90's cells compute the XOR of the two neighboring values at the previous time step, and therefore it always converges to the all-zeros state when the sequence size is a power of 2.}
    \label{fig:rule90_progression}
\end{figure}
\textbf{Figure \ref{fig:rule90_progression}} shows the LWC component for rule 90. As we can see the overall behavior for this rule alternate between two different structures that depends on the sequence length being odd or even. Moreover, since rule 90 computes the eXclusive-OR (XOR) of neighbors, when the sequence length is a power of 2 every state converges to the all-$\square$ self-loop. Because every state converge to the same state the whole network for this rule forms a single LWC component.

\begin{figure}[!htb]
    \centering
    \includegraphics[width=\linewidth]{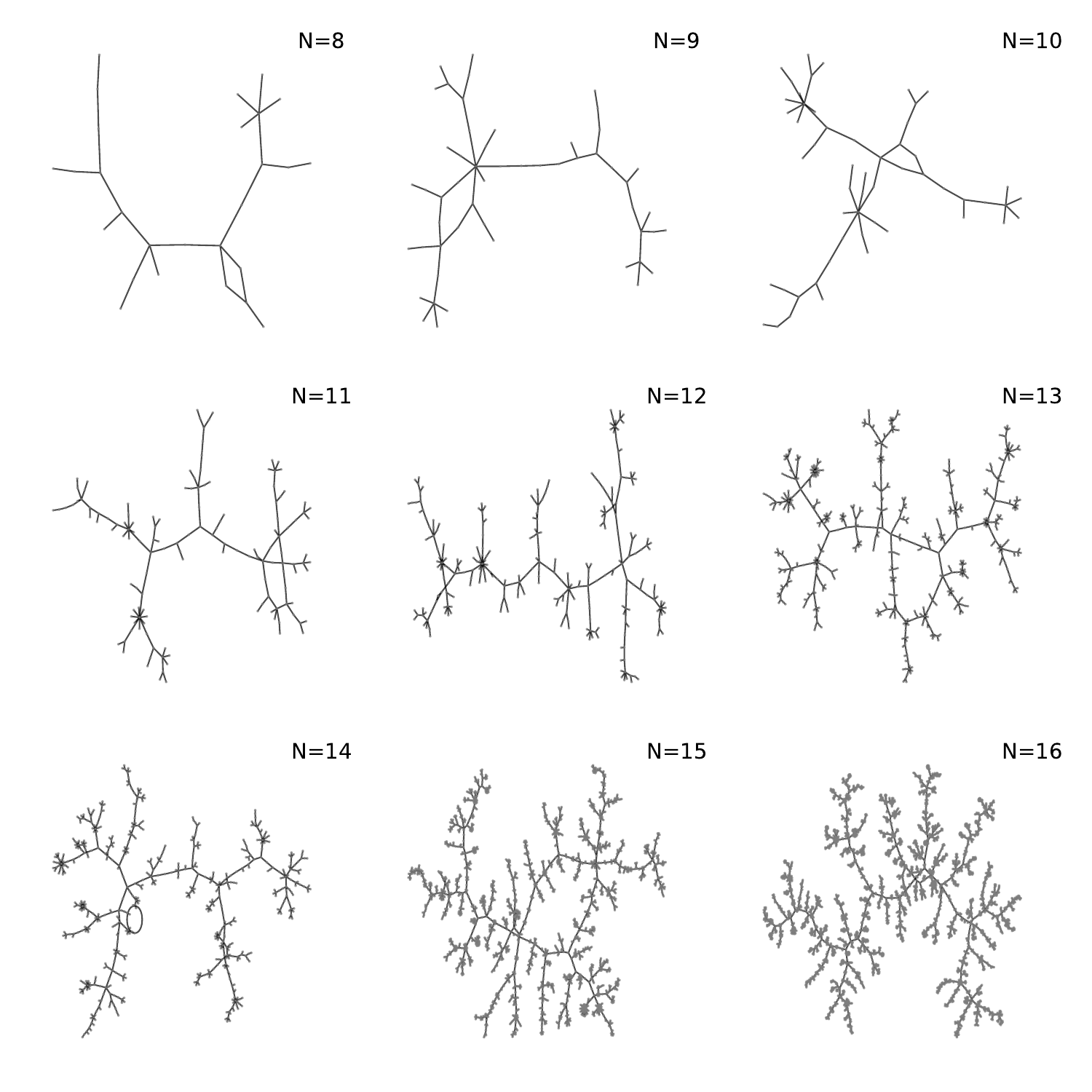}
    \caption{Progression of rule 110 for sequence lengths $N=[8,...,16]$, \textit{only largest} WC component is shown.}
    \label{fig:grid_110}
\end{figure}

While the growth pattern of rule 90 appears to oscillate between two different modes for odd and even sequence lengths, other rules have a more fractal growth. For example, we can see the growth of rule 110 in \textbf{Figure \ref{fig:grid_110}}. Rule 110 is quite interesting because it has been shown that it is capable of universal computation \citep{cook2004universality}.
Despite the intricate branching structure of the LWC component of rule 110 we can see that each network shown contains one central loop that all states in that component converge to.

The existence of an attractor loop in each component is a consequence of the structure of ECA rules. Since ECA rules are deterministic every state (and by extension every necklace) has exactly one outgoing edge. This constraint implies that there \textit{cannot be any edges going away from a loop}, 
\begin{wrapfigure}[7]{r}{0.4\linewidth}
  \centering
  \vspace{-\intextsep}
  \hspace{-\columnsep}
  \includegraphics[width=\linewidth]{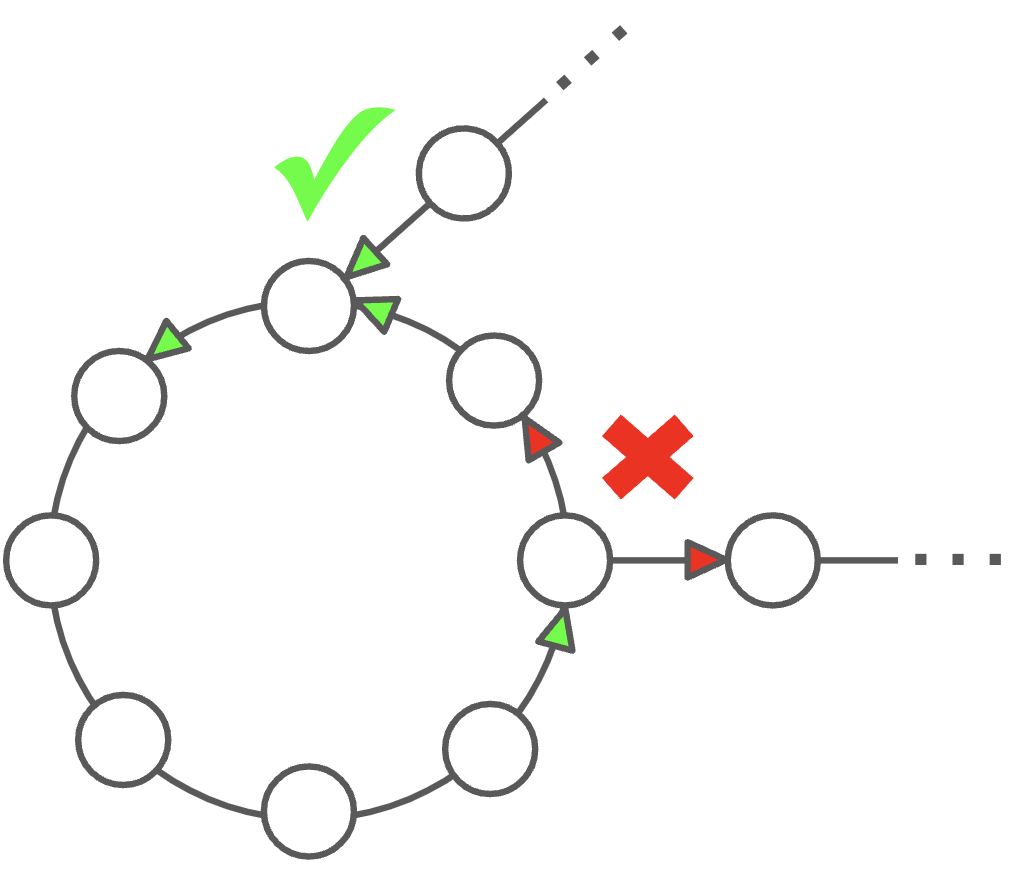}
\end{wrapfigure}
since by definition every node in a loop needs one outgoing connection to the next node in the loop and there is only one outgoing connection available. Furthermore, each node provides exactly one outgoing edge, which implies that there is at least one loop per weakly connected (WC) component. Only $N-1$ edges are needed to connect $N$ nodes into a WC component, therefore the last edge will necessarily connect to a node already in the component. Since belonging to the same WC component means that a path already exists between them, this last edge will necessarily form a loop. These two constraints together imply that every WC component contains \textit{exactly} one loop (a self-loop in the smallest case).

\begin{figure}[!htb]
    \centering
    \includegraphics[width=\linewidth]{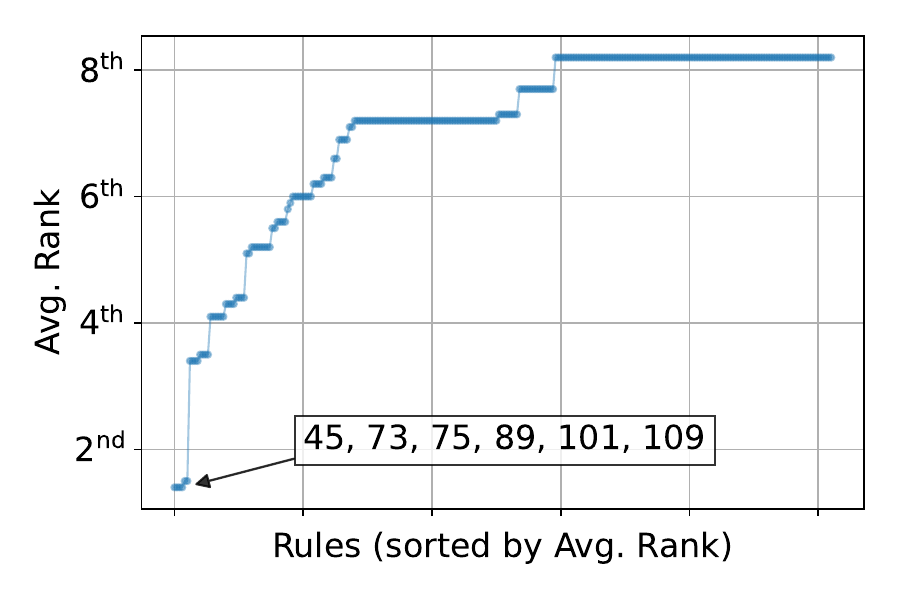}
    \caption{Average rank of ECA rules for values of N from 12 to 21, when sorted by number of unique loop lengths.}
    \label{fig:cycle_ranks}
\end{figure}

Since every WC component contains exactly one loop, it is interesting to further investigate the structure of loops generated by different rules. In particular, we can look at what are the lengths of loops present at the core of each component. While most rules contain many loops of all the same length, some rules contain several loops of varying lengths. If we look at the number of unique loop lengths, a few rules stand above the rest. \textbf{Figure \ref{fig:cycle_ranks}} shows the average rank of each rule when sorted according to number of unique loop lengths, for sequence lengths ranging from $12$ to $21$. Noticeably, a small group of rules (namely 45, 73,75, 89, 101 and 109) has an average rank $\geq 2^{\text{nd}}$, with a large margin above the runner up. 

\begin{figure}[!htb]
    \centering
    \includegraphics[width=\linewidth]{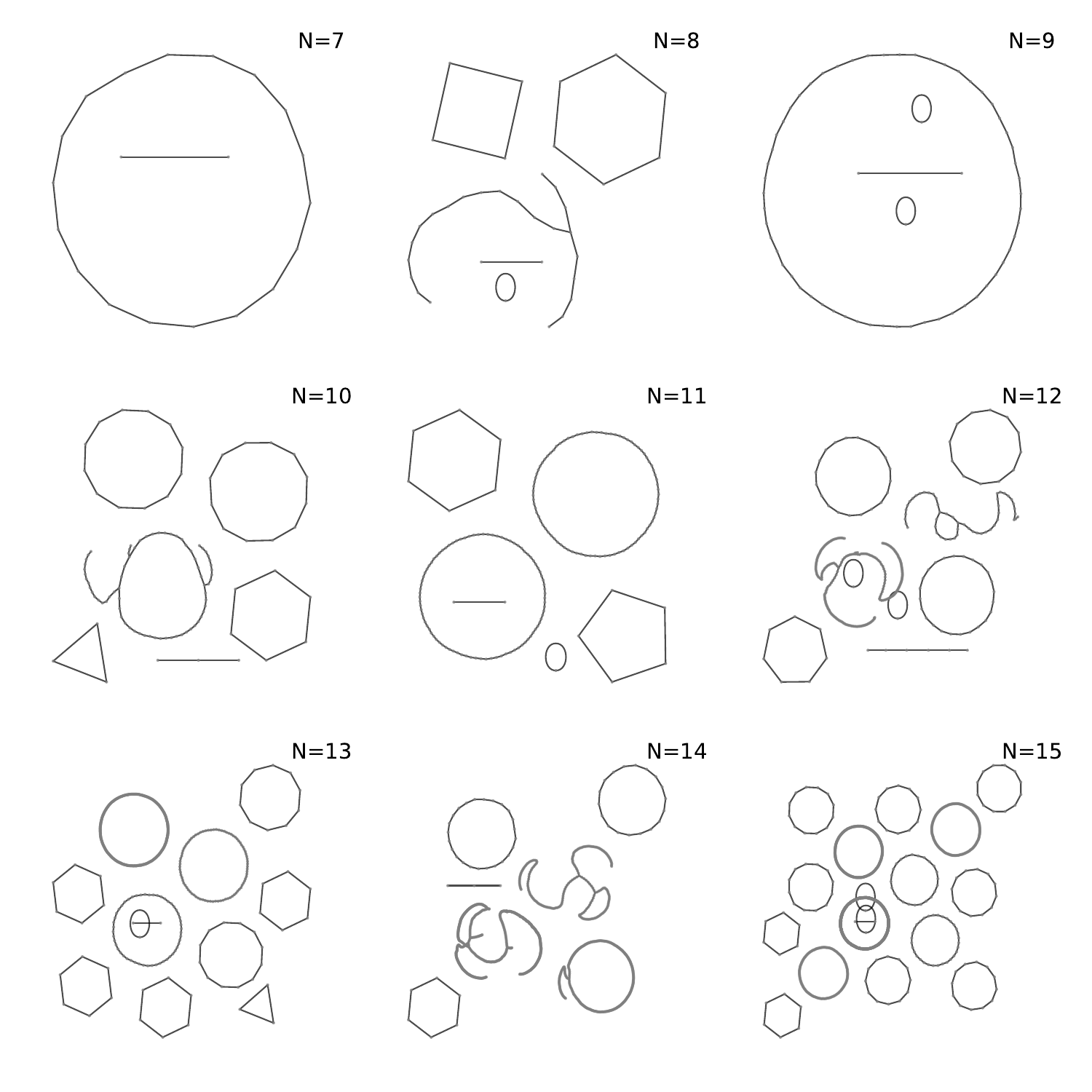}
    \caption{Progression of rule 45 for sequence lengths $N=[7,...,15]$, \textit{all} WC components are shown.}
    \label{fig:grid_45}
\end{figure}

Note that rules 75, 101 and 89 are equivalent to rule 45 by conjugation, reflection and conjugate reflection respectively \cite[Tables of Cellular Automaton Properties p.485-557]{wolfram1986theory}.

\begin{figure}[!htb]
    \centering
    \includegraphics[width=\linewidth]{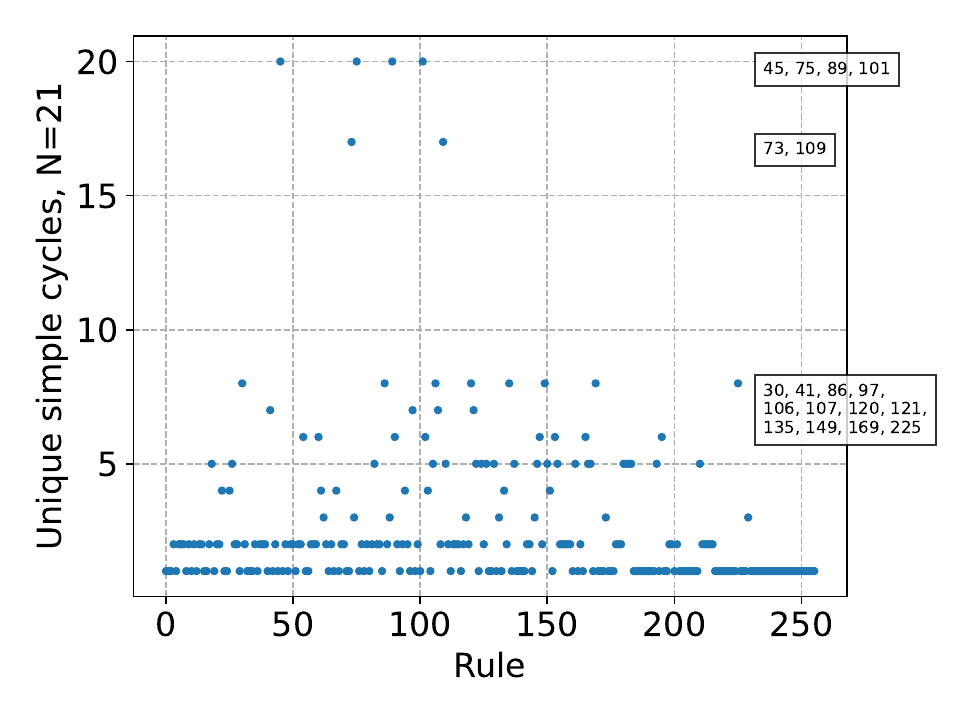}
    \caption{Number of unique cycle lengths across all 256 ECA rules, for $N=21$}
    \label{fig:simple_cycles}
\end{figure}

In \textbf{Figure \ref{fig:grid_45}} we can see \textit{all} the WC components, highlighting how this rule forms a smaller number of components with distinct loop lengths. The total number of unique loop lengths remains fairly small even when the number of nodes in the networks grows. 
For example, \textbf{Figure \ref{fig:simple_cycles}} shows that for $N=21$ (i.e. $\sim 10^5$ nodes) there are only $20$ unique loop lengths. 

\begin{figure}[!hb]
    \centering
    \includegraphics[width=\linewidth]{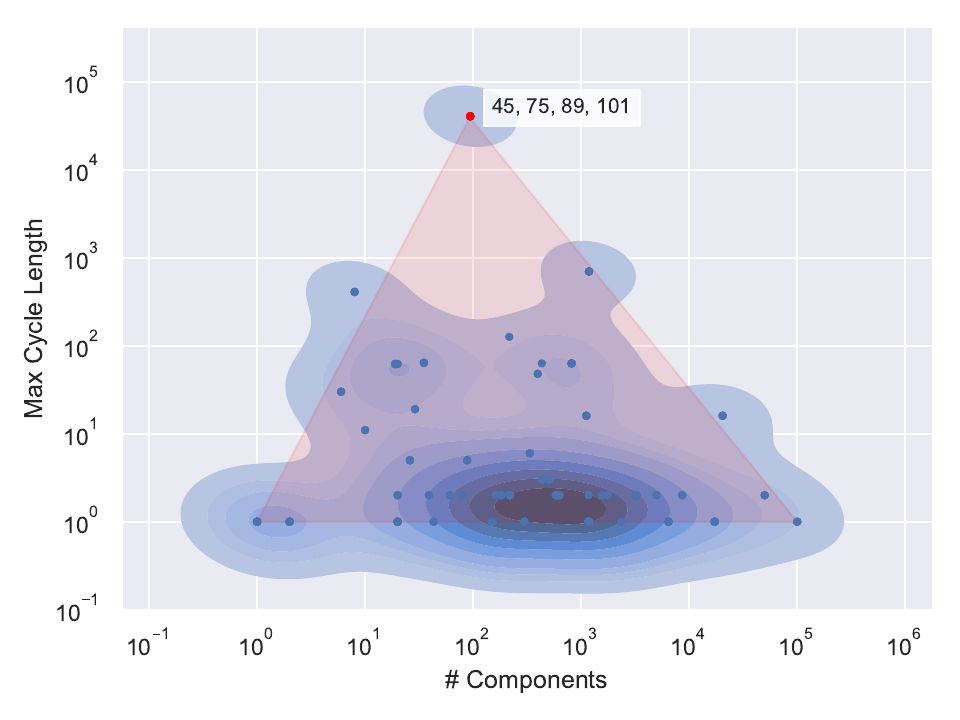}
    \caption{Distribution of ECA rules, $N = 21$, according to max cycle length and \# of weakly connected components: \inlinebox{red} highlights the trade-off between extreme values.}
    \label{fig:max_len_components}
\end{figure}

Notice that since each component can only contain a single loop the total number of loops is equal to the number of components. Also, the more nodes are in a single component, the fewer distinct components a rule can have. The number of nodes per component can grow depending on the length of their core attractor loop (as in rule 45) or as more tree-like structures leading to the attractor loops (as in rule 110). \textbf{Figure \ref{fig:max_len_components}} shows the relationship between max loop length and number of components in each rule. Interestingly, the small group of rules that stands out above the rest is the same as in Figure \ref{fig:simple_cycles}, meaning that those rules both contain the longest loops and the most diverse loop lengths. 

\begin{figure}[!htb]
    \centering
    \includegraphics[width=\linewidth]{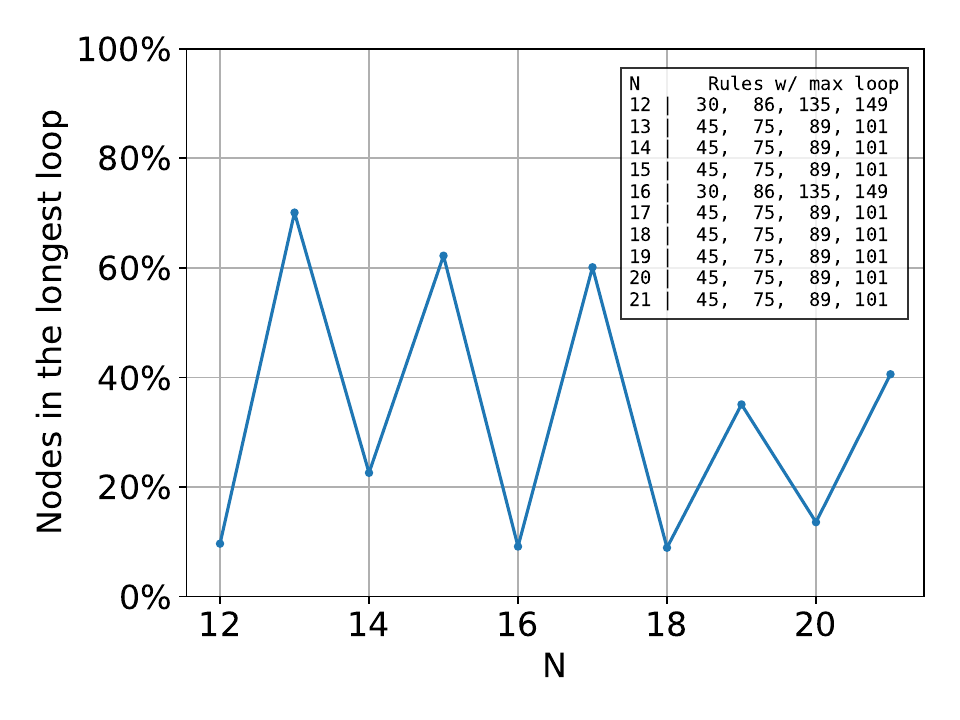}
    \caption{Percentage of all nodes in the necklace networks that belong to the longest loop, for different values of sequence length N (see \citet[p. 232, Fig. 4]{martin1984algebraic} for more information).}
    \label{fig:longest_loops}
\end{figure}

\begin{figure*}[!htb]
    \centering
    \includegraphics[width=\linewidth]{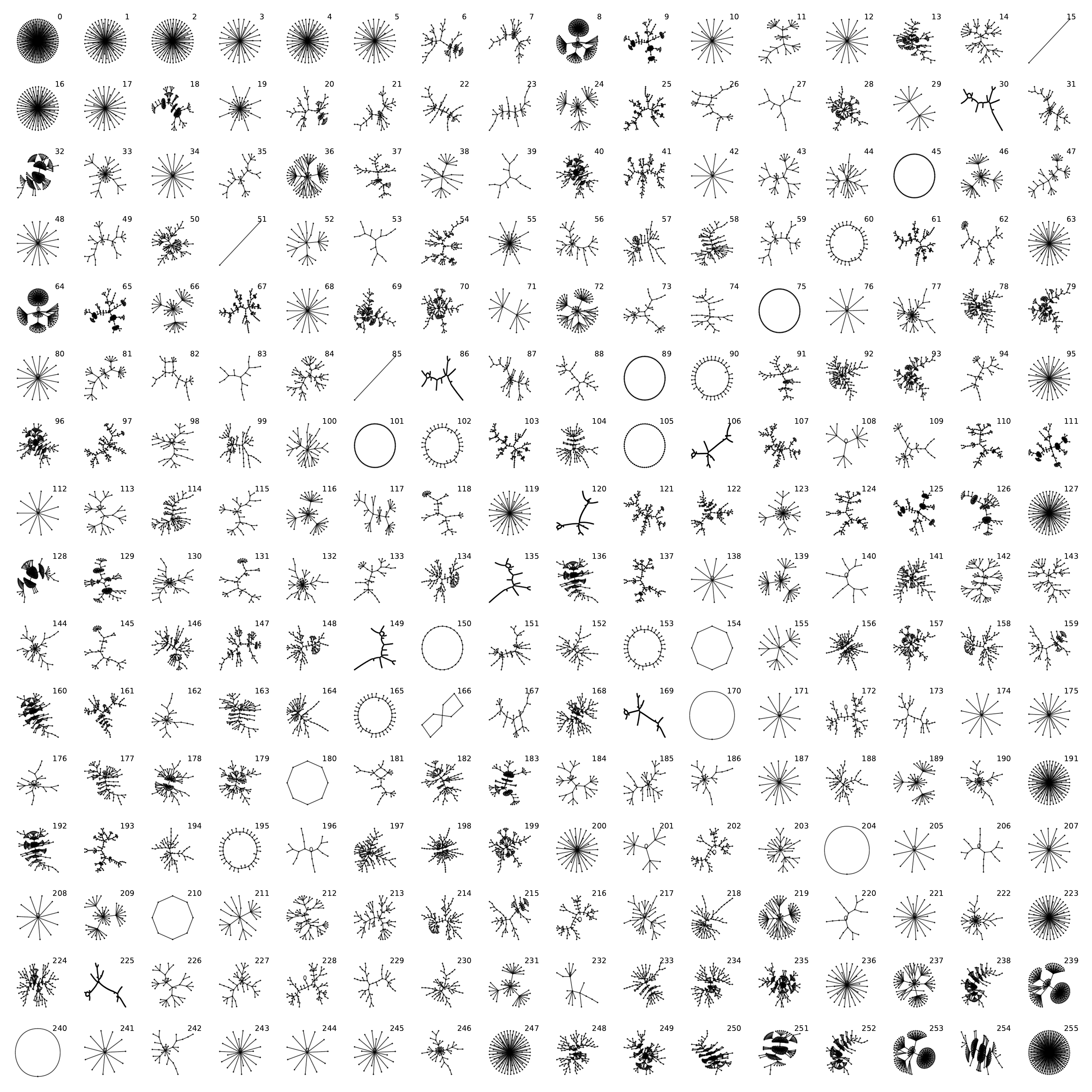}
    \caption{Largest weakly connect component of the transition network induced by the 256 elementary cellular automata on binary 11-necklaces.}
    \label{fig:largest_components}
\end{figure*}

\textbf{Figure \ref{fig:longest_loops}} shows how the longest loop across all rules scales as the sequence length $N$ increases (calculated as the percentage of total nodes that are part of the longest loop, i.e. $100 \times \frac{\max(\text{loop length})}{\# \text{nodes}}$). Figure \ref{fig:longest_loops}-inset shows that rules 30 and 45 have the longest loops, which is supported by previous studies showing them to be computationally chaotic \cite[Hyp. 9]{hudcova2021computational}.
Finally, we show the whole set of 256 ECA rules in \textbf{Figure \ref{fig:largest_components}} for $N=11$.

\section{Conclusion \& Future Work}
The connection between networks and cellular automata has been investigated in several other works \citep{wuensche1993global, kayama2011network} and recently culminated in S. Wolfram's ambitious ``Physics Project"\citep{wolfram2020class}.

Our work has investigated a simple way to reduce the dimensionality of networks generated by ECA rules, namely by exploiting the presence of periodic boundary conditions. We hope our work will help new and seasoned alife practitioners gain a better intuitive understanding of the global structure induced by the application of simple local rules.

From a similar but opposite point of view our work highlights the value of preserving invariants when modeling artificial systems. Namely, the ``flattening" of binary loops into sequences with a fixed beginning and end, compromised their rotational invariance. In the field of deep learning a huge breakthrough followed the popularization of attention mechanism \citep{vaswani2017attention} (i.e., a way to grant permutation equivariance to sequences), perhaps attention could synergize with rotational invariance and unlock new and interesting ideas for ECA in alife.

Because every WC component inevitably converges on one attractor loop (see \citet{wuensche2000basins} for similarly amazing visualizations) any system driven by these rule networks would also eventually converge to a fixed cycle.
In future work we will explore the distance between nodes in different components or across different rules. This will allow us to generate complex and more flexible paths between states, assembling new structure from the building blocks provided by these simple rules, and adding to a growing body of research on using Cellular Automata as a dynamic substrate
 \citep{variengien2021towards, pontes2022unified} or a complex reservoir \citep{yilmaz2014reservoir}, while retaining the simplicity and computational efficiency of the 1-D case.

\section{Acknowledgments}
We thank Michael Drennan for his helpful discussions during the course of this research. This material is based upon work supported by the National Science Foundation under Grants No. 2008413 and 2239691.

\clearpage
\footnotesize
\bibliographystyle{apalike}
\bibliography{references}

\clearpage

\appendix
\section{Overlapping Necklaces}

\begin{figure}[htb]
    \centering
    \includegraphics[width=\linewidth]{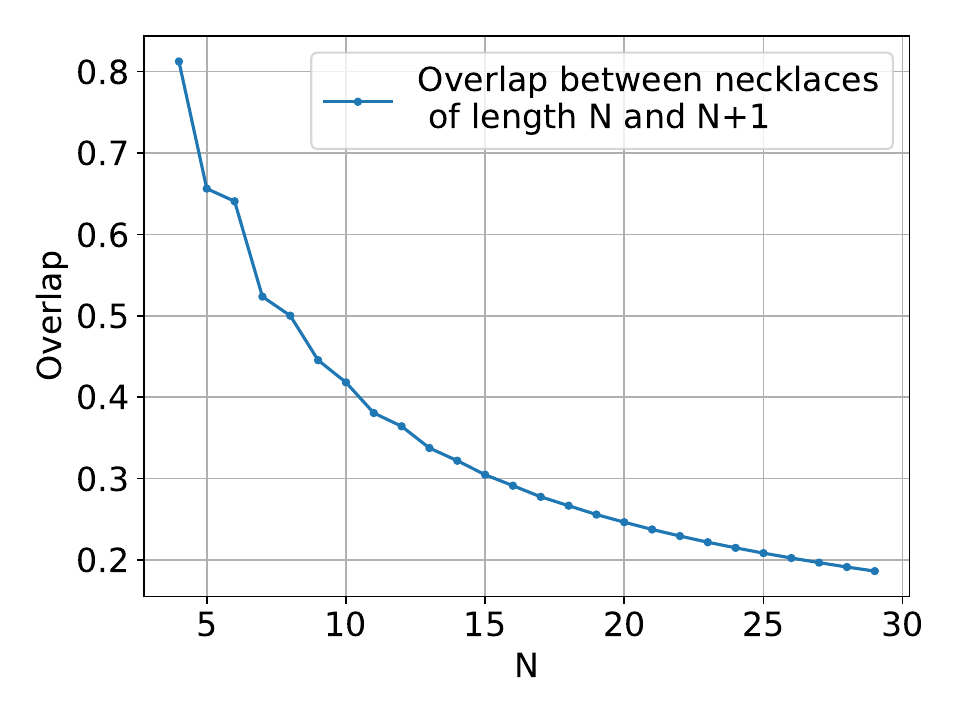}
    \caption{Fraction of representatives that match between consecutive values of N. The trend seem to approach zero as N grows.}
    \label{fig:scaling_overlap}
\end{figure}

\begin{figure}[htbp]
    \centering
        \includegraphics[width=\linewidth]{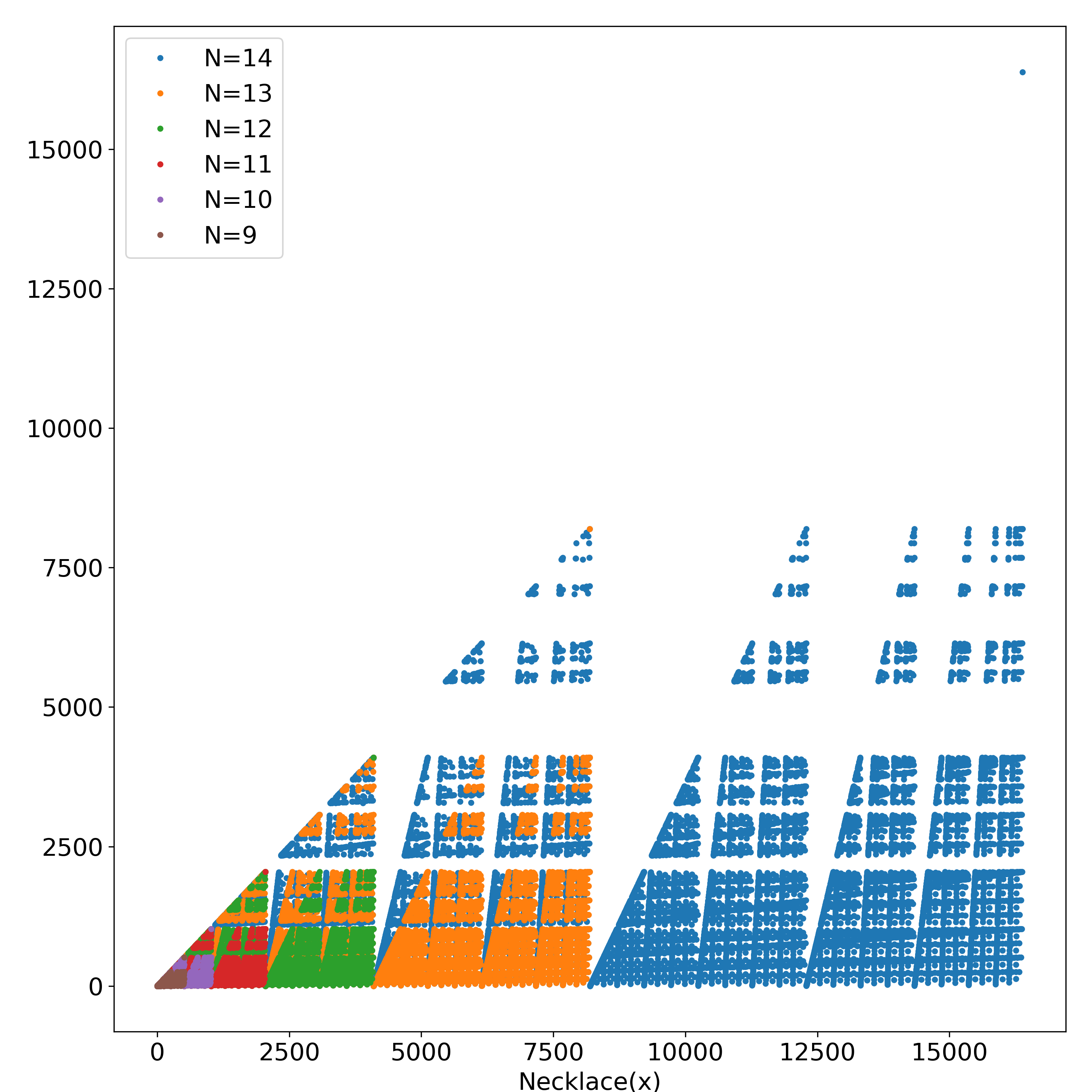}
        \caption{Distributions of necklaces' representatives for consecutive values of N.}
        \label{fig:self_similarity}
\end{figure}

\begin{figure}[htbp]
    \centering
        \includegraphics[width=\linewidth]{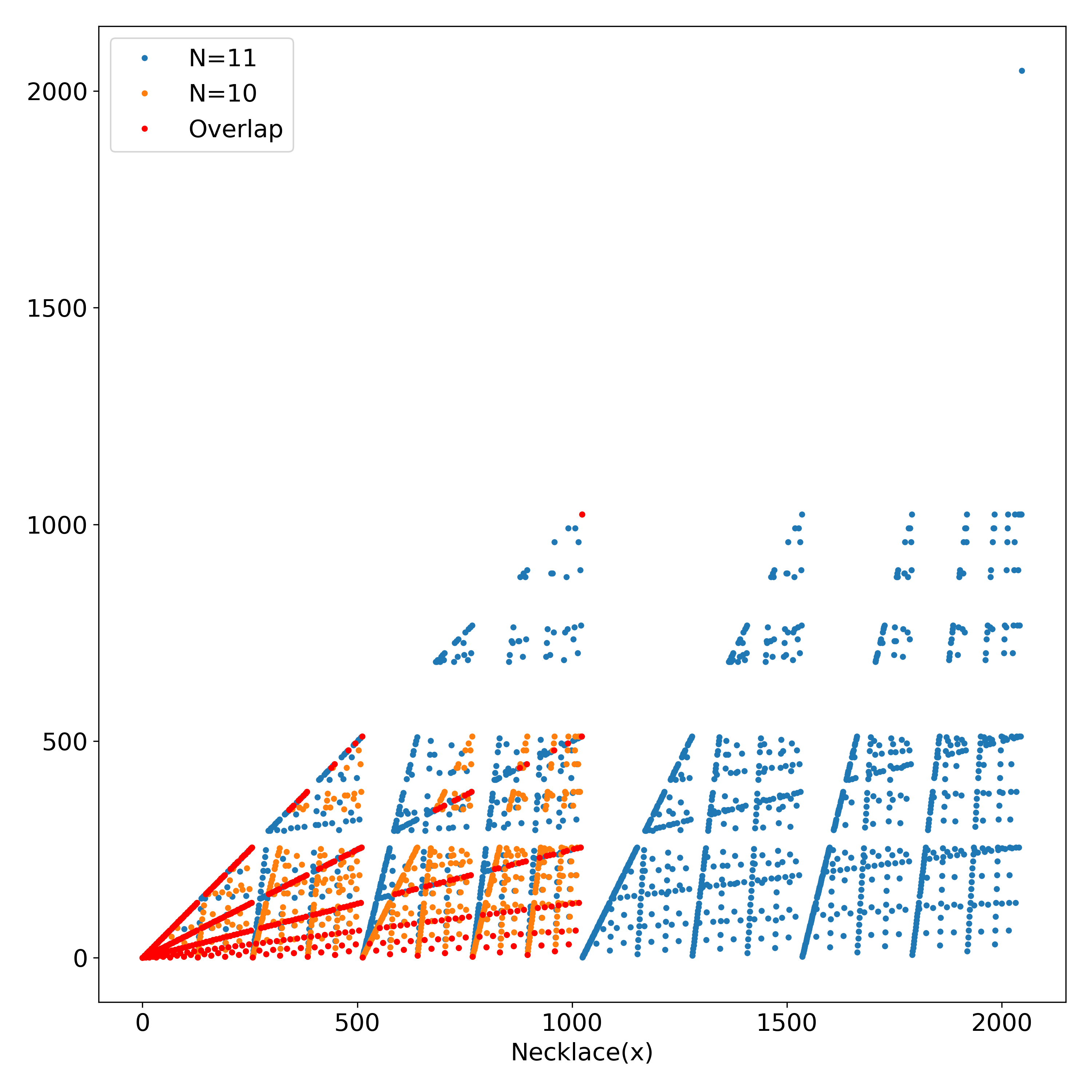}
        \caption{Overlapping between of necklaces' representatives for consecutive values of N.}
        \label{fig:overlap}
\end{figure}

\clearpage
\begin{figure*}
    \centering
    \includegraphics[width=1\linewidth]{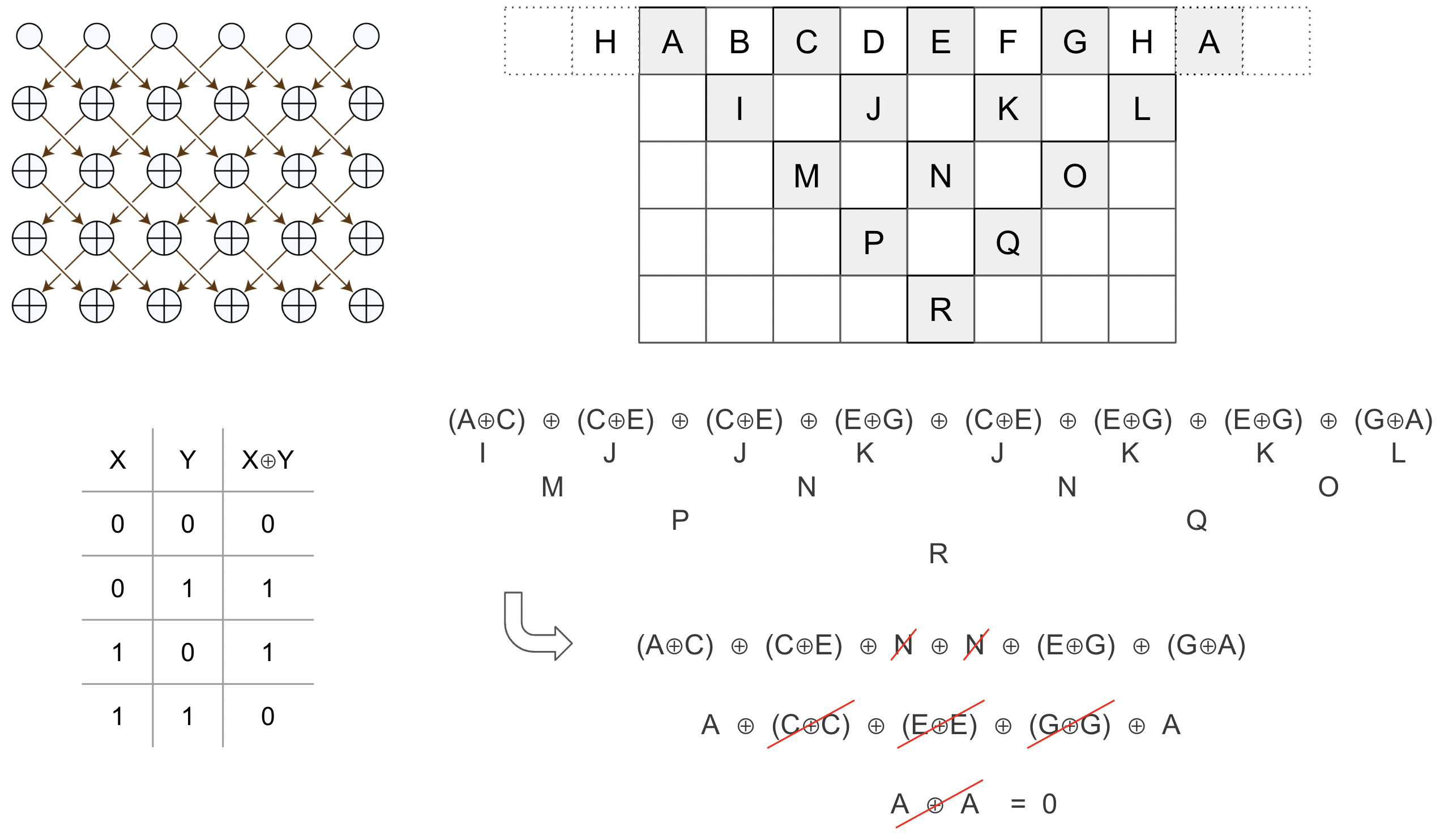}
    \caption{Rule 90 converges to a uniform state all-$\square$ for sequence lengths that are powers of 2. \textit{Left}: Rule 90 computes the $\oplus$ (\textit{bottom}) of the neighbors of each cell (\textit{top}). \textit{Right}: An example of convergence for sequence length $2^3$. We consider the state of cell R in the diagram above, the same reasoning applies for other cells at the same depth. After 4 iterations the state of cell R is determined by the initial state of 5 cells [A,C,E,G,A]. Note that because the starting state was of size $2^3$ and these 5 cells span $2^3+1$ positions, one of these cells (A) is repeated. If we consider the chain of XOR operations that determines the state of cell R we can see that all terms cancel out. Therefore, the state of R doesn't depend on the initial state chosen, instead it will always be $\square$.}
    \label{fig:rule90_xor}
\end{figure*}

\clearpage
\section{All rules}
\begin{figure}[htb]
    \centering
    \includegraphics[width=1\linewidth]{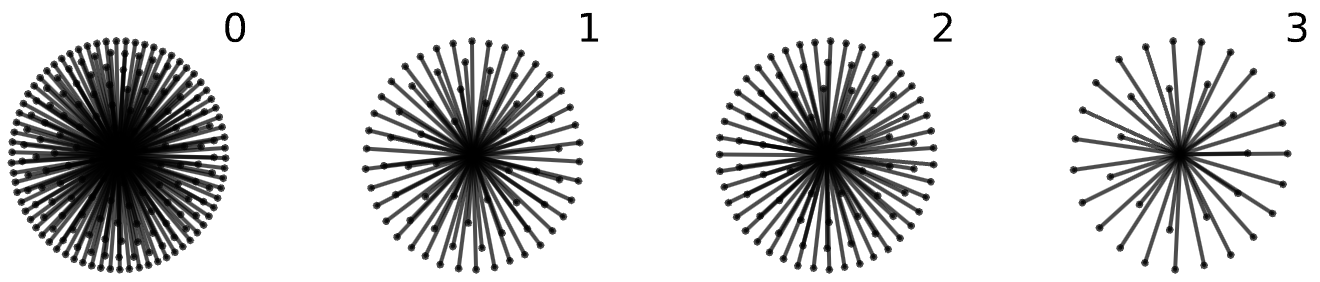}
    \caption{Some rules have states with a large amount of incoming edges. The most extreme examples being rules 0 and 255 that send every sequence into the all zeros/ones sequence respectively.}
    \label{fig:all_11_spiky}
\end{figure}

\begin{figure}[htb]
    \centering
    \begin{subfigure}[t]{0.65\linewidth}
        \includegraphics[width=\linewidth]{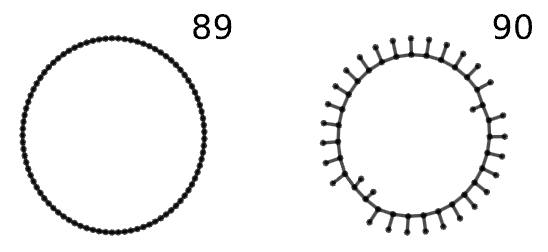}
        \caption{Several rules form long loops of either single states (\textit{left}) or sprouts (\textit{right}).}
        \label{fig:all_11_interesting}
    \end{subfigure}
    \hfill
    \begin{subfigure}[t]{0.28\linewidth}
        \includegraphics[width=\linewidth]{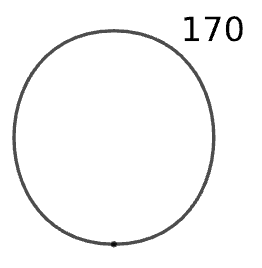}
    \caption{Rules that only shift contain only self-loops}
    \label{fig:all_11_boring}
    \end{subfigure}
    \caption{Three kinds of loops.}
    \label{fig:all_11_loops}
\end{figure}

\begin{figure}[htb]
    \centering
    \includegraphics[width=\linewidth]{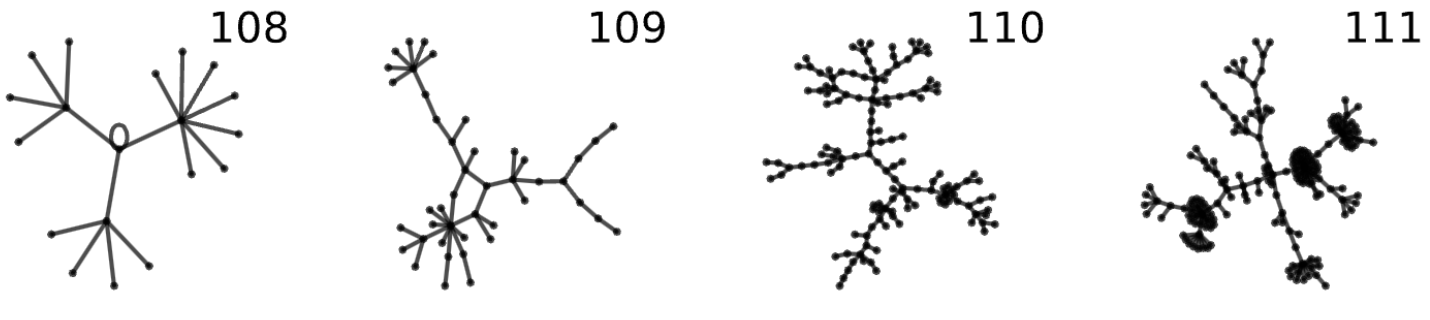}
    \caption{Trees and large loops belong to separate components. Note that the fewer the nodes in the LWC component, the more distinct components the rule induces in the necklace network.}
    \label{fig:all_11_trees}
\end{figure}

The LWC components in Figure \ref{fig:largest_components} roughly fall into 3 categories shown in Figures SI.\ref{fig:all_11_spiky}, SI.\ref{fig:all_11_loops} and SI.\ref{fig:all_11_trees}.

\begin{figure}[!hb]
    \centering
    \includegraphics[width=0.85\linewidth]{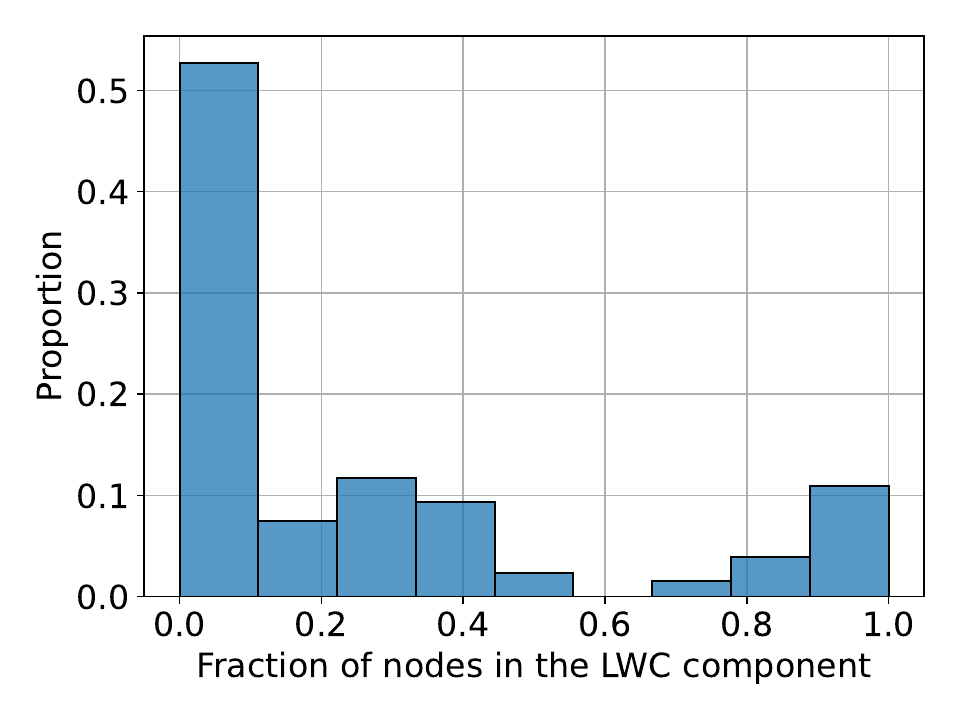}
    \caption{Histogram of values in Figure \ref{fig:LWC_coverage_21}}
    \label{fig:LWC_distribution_21}
\end{figure}

\begin{figure}[htb]
    \centering
    \includegraphics[width=\linewidth]{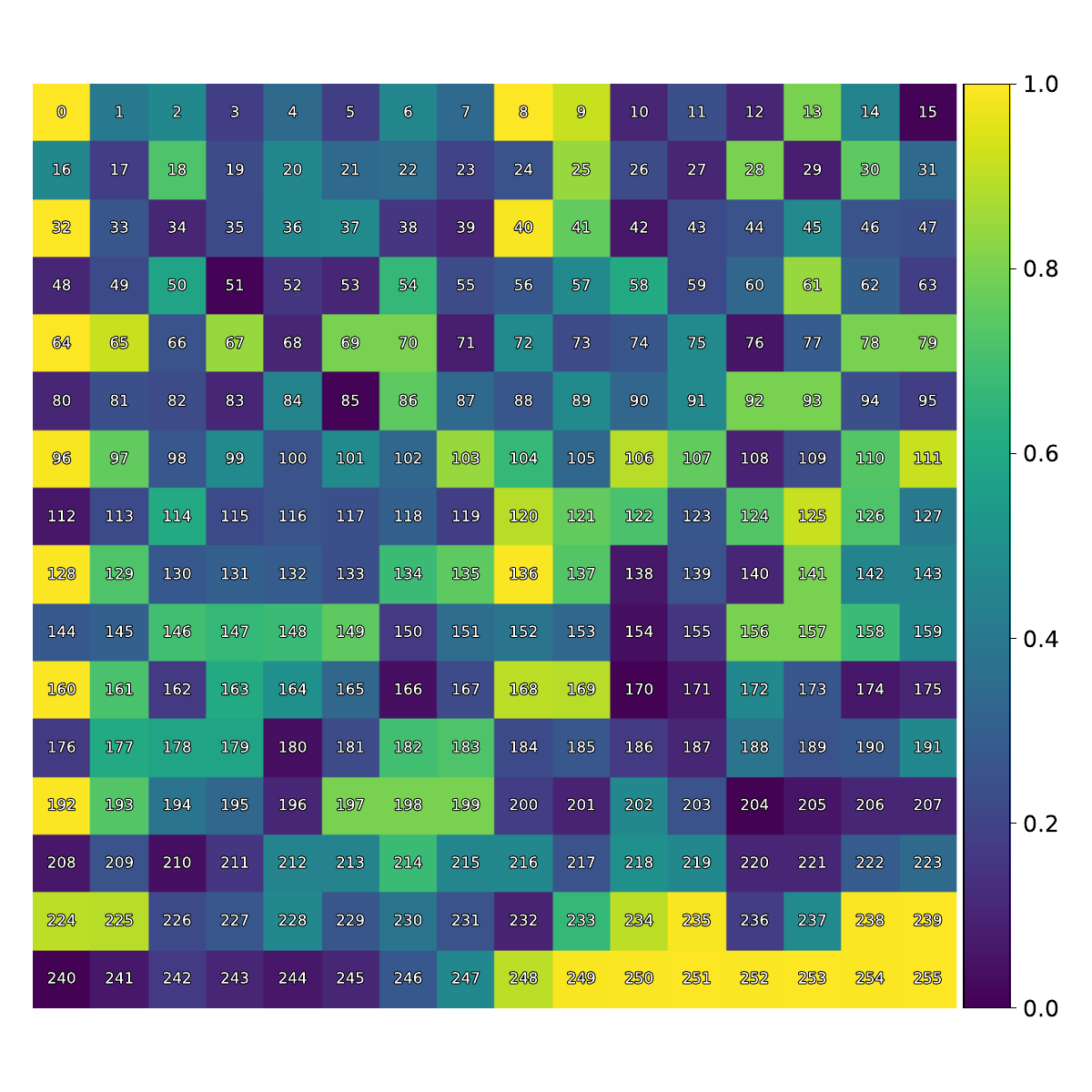}
    \caption{Fraction of nodes in the LWC component for each rule shown in Figure \ref{fig:largest_components} for $N=11$.}
    \label{fig:LWC_coverage_11}
\end{figure}

\begin{figure}[htb]
    \centering
    \includegraphics[width=\linewidth]{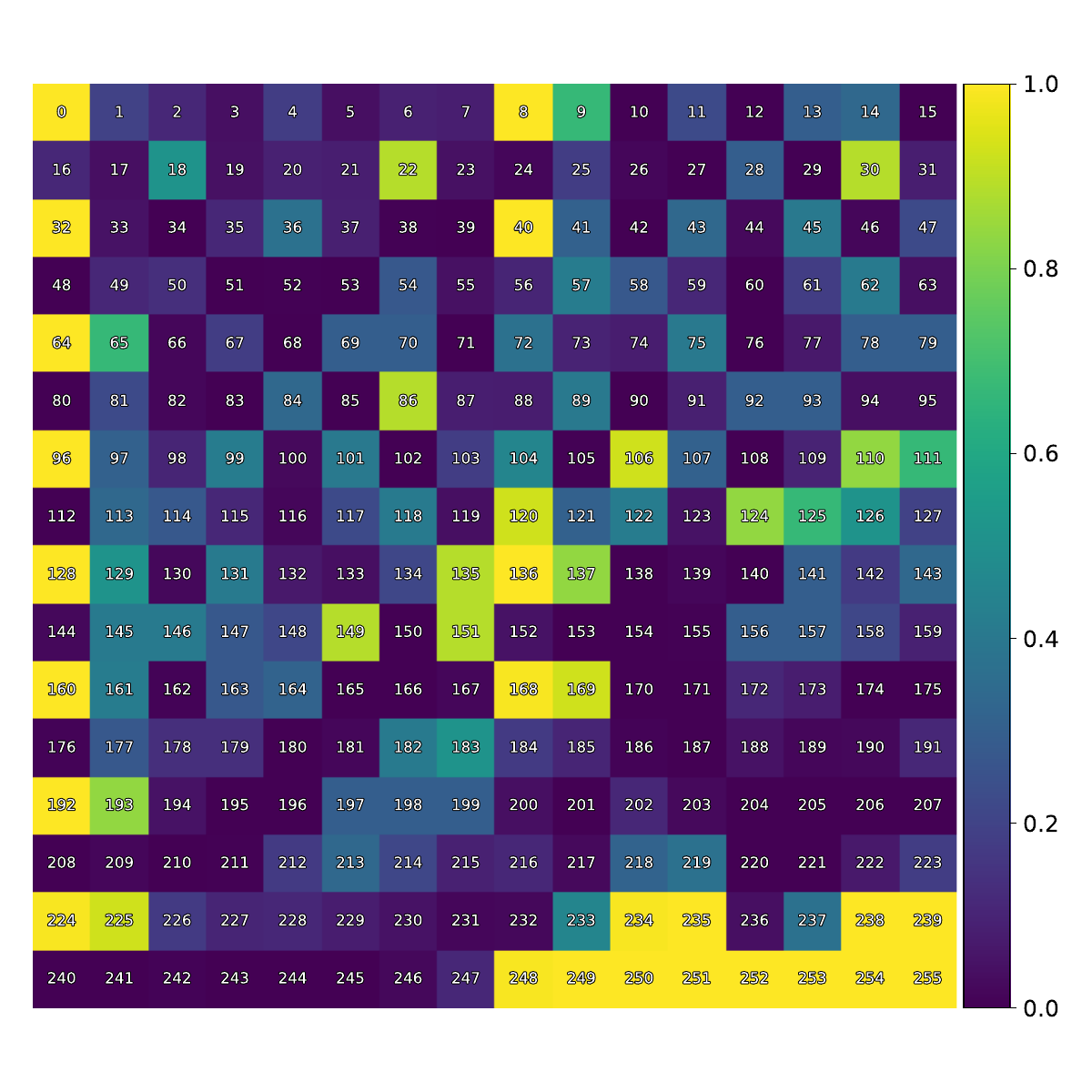}
    \caption{Fraction of nodes in the LWC component for each ECA rule for $N=21$. As the value of N grows the distribution of values tends to accumulate towards 0 and 1. See Figure \ref{fig:LWC_distribution_21}.}
    \label{fig:LWC_coverage_21}
\end{figure}

\end{document}